\documentclass[aps,prl,twocolumn,superscriptaddress,showpacs,preprintnumbers,amsmath,amssymb,dvips]{revtex4}

\newif\ifpdf

\ifx\pdfoutput\undefined
  \pdffalse 
  \else
  \pdfoutput=1 
  \pdftrue \fi

  \ifpdf \usepackage[pdftex]{graphicx}
  \graphicspath{{PDF/}{:PDF:}}
  \DeclareGraphicsExtensions{.jpg,.pdf,.mps,.png} \else
  \usepackage{graphicx} \graphicspath{{EPS/}{:EPS:}}
  \DeclareGraphicsExtensions{.eps,.ps,.bmp,.tif,.tiff,.tga} \fi

\usepackage{mathptmx}

\usepackage{dcolumn}

\usepackage{verbatim}

\usepackage{subfigure}

\usepackage{rcs}
\rcsAuthor{scentoni}{Scott A Centoni}

\usepackage[squaren,textstyle]{SIunits}
\addunit{\eV}{\electronvolt}
\addunit{\Rydberg}{Ry}
\addunit{\Ao}{\angstrom}
\addunit{\wavenumber}{\reciprocal{\centi\meter}}

\usepackage{hyphenat}

\usepackage{keyval}

\usepackage{braket}

\usepackage[T1]{fontenc}

\setkeys{Gin}
{keepaspectratio=true,
totalheight=0.25\textheight,
width=0.99\columnwidth}

\newcolumntype{.}{D{.}{.}{-1}}
\newcommand{\ccol}[1]{\multicolumn{1}{c}{#1}}

\newcommand{\mycap}[2]{\caption[#1]{\textbf{#1} #2} }

\newcommand{\proper}[1]{\textsc{\small #1}}

\usepackage{amsfonts}

\newcommand{\nablash}{\nabla{\kern -.74 em
   \raise 1.5 true pt\hbox{{\bf/}}}\kern +.1 em}

\newcommand{\eva}[1]{  }

\newcommand{\deriv}[2]{\frac{\partial #1}{\partial #2}}

\newcommand{\textderiv}[2]{\textfrac{\partial #1}{\partial #2}}

\DeclareMathAlphabet{\mathsfsl}{OT1}{cmss}{m}{sl}

\newcommand{\mathvec}[1]{\mathbf{#1}}
\newcommand{\mathmat}[1]{\mathsf{#1}}

\newcommand{\electron}{{\text{e}^-}}

\newcommand{\level}{\varepsilon}
\newcommand{\fermi}{_{\text{F}}}

\newcommand{\valence}{_{\text{v}}}
\newcommand{\conduction}{_{\text{c}}}
\newcommand{\electrons}{_{\text{e}}}
\newcommand{\holes}{_{\text{h}}}

\newcommand{\intrinsic}{_{\text{i}}}

\newcommand{\equilibrium}{^{\text{eq}}}
\newcommand{\equilibriumzero}{^{\text{eq}0}}
\newcommand{\eq}{\equilibrium}
\newcommand{\eqz}{\equilibriumzero}

\newcommand{\formation}{^{\text{f}}}
\newcommand{\relaxation}{^{\text{rel}}}
\newcommand{\migration}{^{\text{m}}}
\newcommand{\diffusion}{^{\text{D}}}

\usepackage{color}

\usepackage{chemsym}

\newcommand{\half}{\frac{1}{2}}

\newcommand{\textfrac}[2]{\ensuremath{ #1/#2 } }

\newcommand{\xx}{\ensuremath{\!\times\!}}

\newcommand{\E}[1]{\ensuremath{\times\!10^{#1}}}

\newcommand{\expnkt}[1]{\ensuremath{\exp{\left(-\frac{#1}{\kB
          T}\right)}}}

\newcommand{\etal}{\emph{et al.}}
\newcommand{\ie}{\emph{i.e.}}
\newcommand{\eg}{\emph{e.g.}}

\newcommand{\kB}{k_B}

\newcommand{\auspices}{This work was performed under the auspices
  of the U.~S. Department of Energy by the University of
  California, Lawrence Livermore National Laboratory under
  Contract No. W-7405-Eng-48.  Funding was provided by the
  Department of Energy Office of Basic Energy Sciences.}

\newcommand{\vol}{v_\Si}
\renewcommand{\diffusion}{^{\text{d}}}

\RCS $Revision: 1.52 $
\RCS $Date: 2002/10/24 00:11:11 $
\RCS $State: Exp $

\begin{document}

 \preprint{APS/123-QED}

\title{First-principles calculation of \\
   intrinsic defect formation volumes in silicon}

\author{Scott A. Centoni}
\email{scentoni@email.sjsu.edu}
\affiliation{
  Department of Materials Engineering,
  San Jose State University,
  San Jose, California, 95192}
\affiliation{
  Lawrence Livermore National Laboratory,
  Livermore, California, 94550}

\author{Babak Sadigh}
\author{George H. Gilmer}
\affiliation{
  Lawrence Livermore National Laboratory,
  Livermore, California, 94550}

\author{Thomas J. Lenosky}
\affiliation{
  Ohio State University,
  Columbus, Ohio 43210}

\author{Tom{\'a}s D{\'\i}az de la Rubia}
\affiliation{
  Lawrence Livermore National Laboratory,
  Livermore, California, 94550}

\author{Charles B. Musgrave}
\affiliation{
  Department of Materials Science and Engineering,
  Stanford University,
  Stanford, California, 94305}

\affiliation{
  Department of Chemical Engineering,
  Stanford University,
  Stanford, California, 94305}

\date{\small Revision\RCSRevision, \RCSDate, \RCSTime\ UTC}

\begin{abstract}
  We present an extensive first-principles study of the pressure
  dependence of the formation enthalpies of all the know vacancy and
  self-interstitial configurations in silicon, in each charge state
  from $-2$ through $+2$.  The neutral vacancy is found to have a
  formation volume that varies markedly with pressure, leading to a
  remarkably large negative value ($-0.68$ atomic volumes) for the
  zero-pressure formation volume of a Frenkel pair ($V + I$).  The
  interaction of volume and charge was examined, leading to
  pressure--Fermi level stability diagrams of the defects.  Finally,
  we quantify the anisotropic nature of the lattice relaxation around
  the neutral defects.
\end{abstract}

\maketitle

\section{Introduction}
\label{sec:introduction}
Nearly perfect crystals of silicon are of great technological
importance, yet silicon self-diffusion is still not completely
understood.  Unlike the situation in metals, the equilibrium
concentrations of vacancies and self-interstitials in Si are
believed to comparable, and very low, making detection of them
problematic.  Experimental data is fragmentary, and simulations do
not all agree.  Controversy remains over the relative importance
of vacancies and interstitials to self-diffusion at different
temperatures and the relative magnitudes of the migration enthalpy
to the formation enthalpy of these defects \cite{Ural99b,Bracht98}.
We set aside the possibility of diffusion by a direct exchange
mechanism \cite{Pandey86} due to the low prefactor that has been
calculated \cite{Pandey90}.  Then the self-diffusivity of silicon
is the sum of the diffusion of Si due to vacancies and due to
interstitials,
\begin{equation}
D_\Si = c_V D_V + c_I D_I
\end{equation}
where the atomic fraction $c_X = C_X/C_\Si$ and $C_\Si =
\unit{5.00\E{22}}{\rpcubic{\centi\meter}}$.  The contribution of
vacancies to Si diffusion is proportional to the concentration of
vacancies $C_V$ and the diffusivity of vacancies $D_V$, and likewise
for interstitials.  The equilibrium concentration of a defect $X$ is
$C\eq_X = C_\Si \exp{\left(-g\formation_X/\kB T\right)}$, where
$g\formation_X$ is the Gibbs free energy of formation of one defect.
The diffusivity of a vacancy can be written as $D_V =
(\zeta/6)\lambda^2 \nu_0 \exp{\left(-g\migration_V/\kB T\right)}$,
where $\zeta$ is the coordination number, $\lambda$ is the bond
length, $\nu_0$ is an attempt frequency, and $g\migration_V$ is the
Gibbs free energy for the vacancy to exchange with one of its
neighbors.  The diffusivity of self-interstitials can be written
similarly, but with a different geometric factor.  Of course, $G = H -
T S$, and if the entropy and enthalpy are assumed to be constant with
respect to temperature, entropy may be combined with the
pre-exponential factor, leaving only the enthalpy as a model parameter
in the exponent.

Recent isotope tracer experiments \cite{Bracht98} fit equilibrium
silicon self-diffusivity to a single Arrhenius term
\begin{equation}
\label{eq:Bracht}
  D\eq_\Si = D\eqz_\Si \expnkt{h\diffusion_\Si}
\end{equation}
with $D\eqz_\Si =
\unit{530_{-170}^{+250}}{\centi\meter\squared\per\second}$ and
$h\diffusion_\Si = \unit{4.75 \pm 0.04}{\eV}$, suggesting that either
vacancies or self-interstitials dominate self-diffusion over the
entire temperature range studied---or instead that they switch over
from one to the other but with similar values of $h\diffusion_\Si$.
Of course, vacancies and self-interstitials diffuse by exchanging with
lattice atoms, and thus cannot be isotopically tagged.

The concentrations $C$ and diffusivities $D$ of vacancies in silicon
are difficult to measure separately with any accuracy, and likewise
with self-interstitials.  Estimated equilibrium transport capacities
$c\eq_X D_X = d^0_X \exp{\left( -\left( h\formation_X + h\migration_X
    \right) / \kB T \right)}$ have been derived from experimental
studies of metal diffusion in silicon, the most recent of which report
values of $h\formation_V + h\migration_V$ ranging from
\unit{4.03}{\eV} to \unit{4.14}{\eV}, and $h\formation_I +
h\migration_I$ from \unit{4.84}{\eV} to \unit{4.95}{\eV}
\cite{Tan85,Bracht98}.  On the other hand, the latest published work
utilizing dopant diffusion arrives at an estimate of $h\diffusion =
h\formation_V + h\migration_V = \unit{4.86}{\eV}$ and $h\diffusion =
h\formation_I + h\migration_I = \unit{4.68}{\eV}$ \cite{Ural99b},
contrary to the long-held assumption that $h\diffusion_I >
h\diffusion_V$.  Ion implantation, thermal oxidation, and nitridation
increase the concentration of intrinsic defects above their
equilibrium concentrations ($C_X \gg C\eq_X$), increasing the self-
and dopant diffusivity; however, measuring activation enthalpies from
such experiments requires assumptions about traps and other
simplifications to differentiate between interstitial and vacancy
mechanisms.

First-principles methods can be used to separately calculate formation
and migration enthalpies, among other quantities.  The enthalpies can
be obtained from the energies calculated at different volumes as $H =
E + P V + q \level\fermi$, where $\level\fermi$ is the Fermi level and
$q$ is the charge of the defect in electron units.  Since $V =
\textderiv{H}{P}$, the formation volume $v\formation_X$ tells us how
the formation enthalpy $h\formation_X$, and thus the equilibrium
concentration varies with pressure.  Similarly, $v\migration_X$ tells
us how much pressure enhances or retards the migration of a defect
once it has entered the lattice.  This is of technological interest
because of the large stresses and strains that exist near the surfaces
and interfaces of silicon-based integrated circuits (\eg{}
heteroepitaxial growth of Si upon
$\text{Si}_{1-x-y}\text{Ge}_{x}\text{C}_{y}$ or dielectric
substrates), which may have a strong effect on concentrations and
diffusion of intrinsic defects.

\section{Methods}
\label{sec:methods}
Total energies were calculated using density functional theory
(DFT), as implemented in the code \proper{VASP}
\cite{Kresse93,Kresse94,Kresse96a,Kresse96b}.  All calculations
were performed with the PW91 exchange--correlation functional.
Ion cores were represented with Vanderbilt ultra-soft
pseudopotentials, allowing plane-wave energy cutoffs of
$\unit{11}{\Rydberg} = \unit{150}{\eV}$.  The Brillouin zone was
sampled with $\mathvec{k}$ points equivalent to a $4 \xx 4 \xx 4$
Monkhorst--Pack mesh in a conventional cubic (8-atom) cell.
Periodic boundary conditions were used with primitive (2-atom)
silicon cells for calculation of pure silicon and larger (mostly
128- or 256-site) supercells for calculations involving defects.

Our reference point was perfect silicon in the diamond cubic
structure.  The total energies at different volumes were fit to a
Birch--Murnaghan equation of state \cite{Murnaghan44}.  In our
calculations, silicon had a cohesive energy of \unit{4.53}{\eV}
and equilibrium atomic volume $\vol = \unit{20.34}{\Ao\cubed}$.
The bulk modulus at $P = 0$ was $B_0 = \unit{0.88}{\mega\bbar}$
and its pressure derivative was $B_0' = 4.02$.

\begin{figure}[tbp]
  \centering
  \subfigure[$V_L$]{\label{fig:VL}\includegraphics[width=0.15\textwidth]{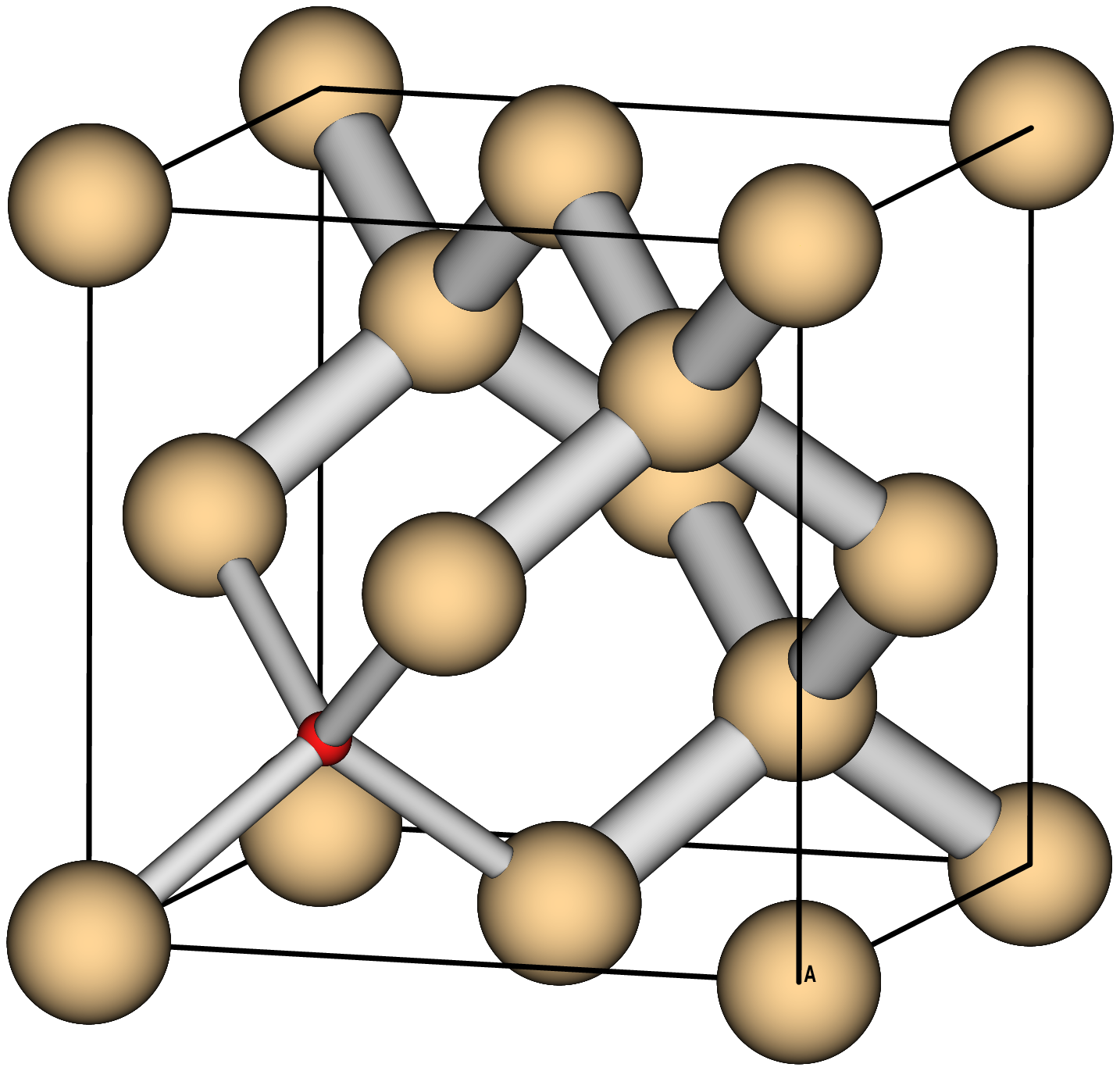}}
  \subfigure[Si]{\label{fig:Si}\includegraphics[width=0.15\textwidth]{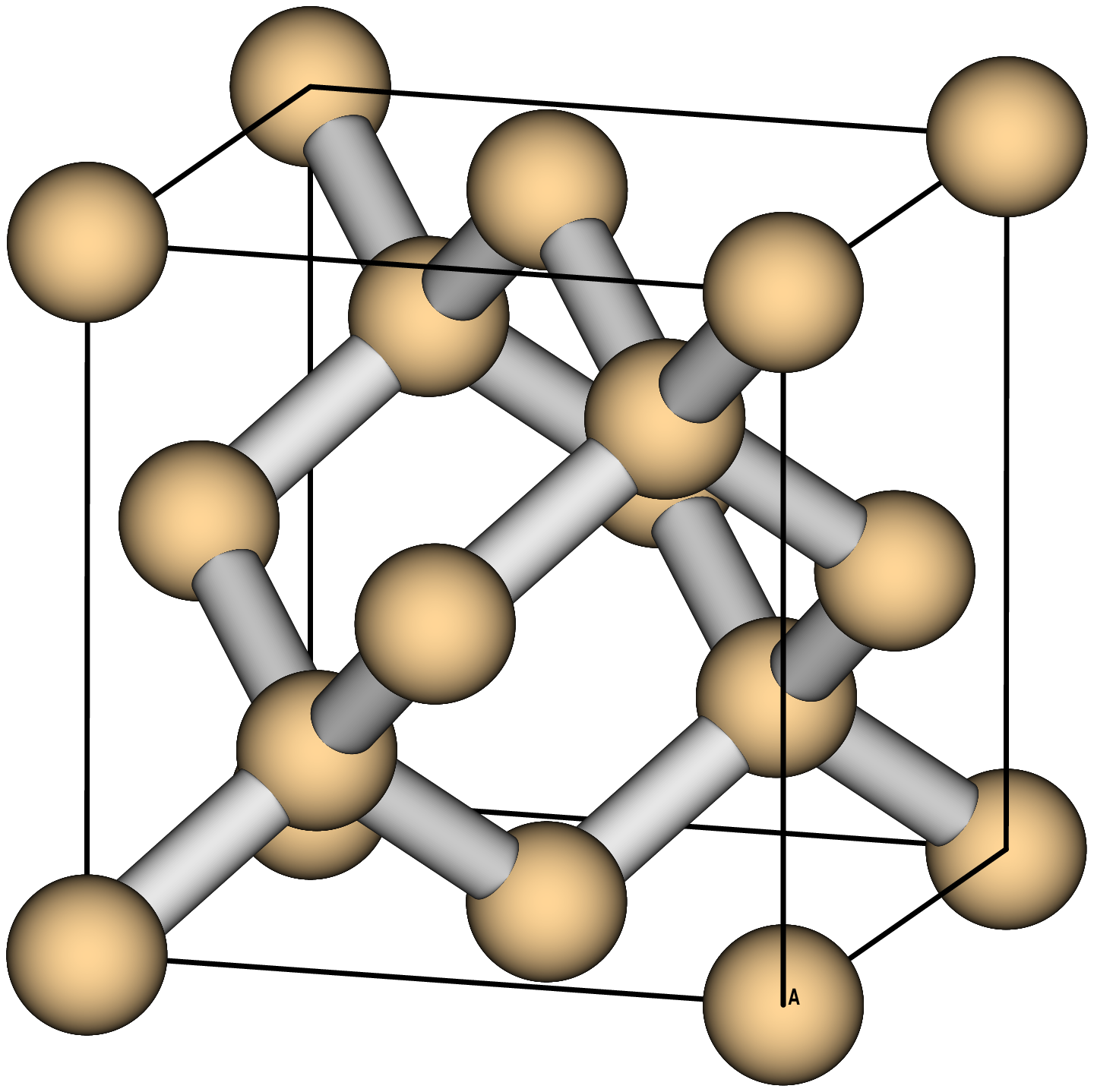}}
  \subfigure[$I_{\braket{110}}$]{\label{fig:IX}\includegraphics[width=0.15\textwidth]{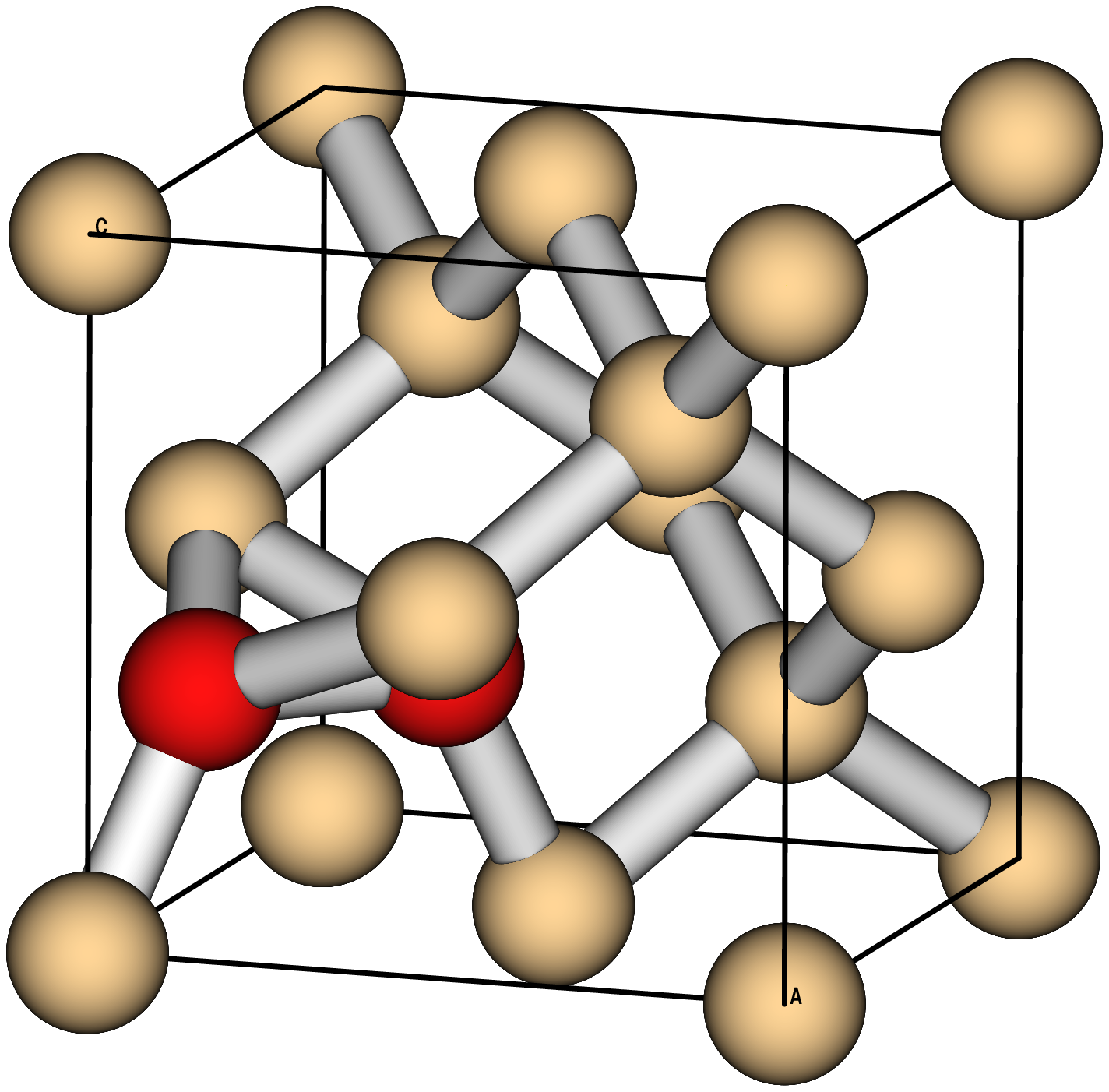}}\\
  \subfigure[$V_B$]{\label{fig:VB}\includegraphics[width=0.15\textwidth]{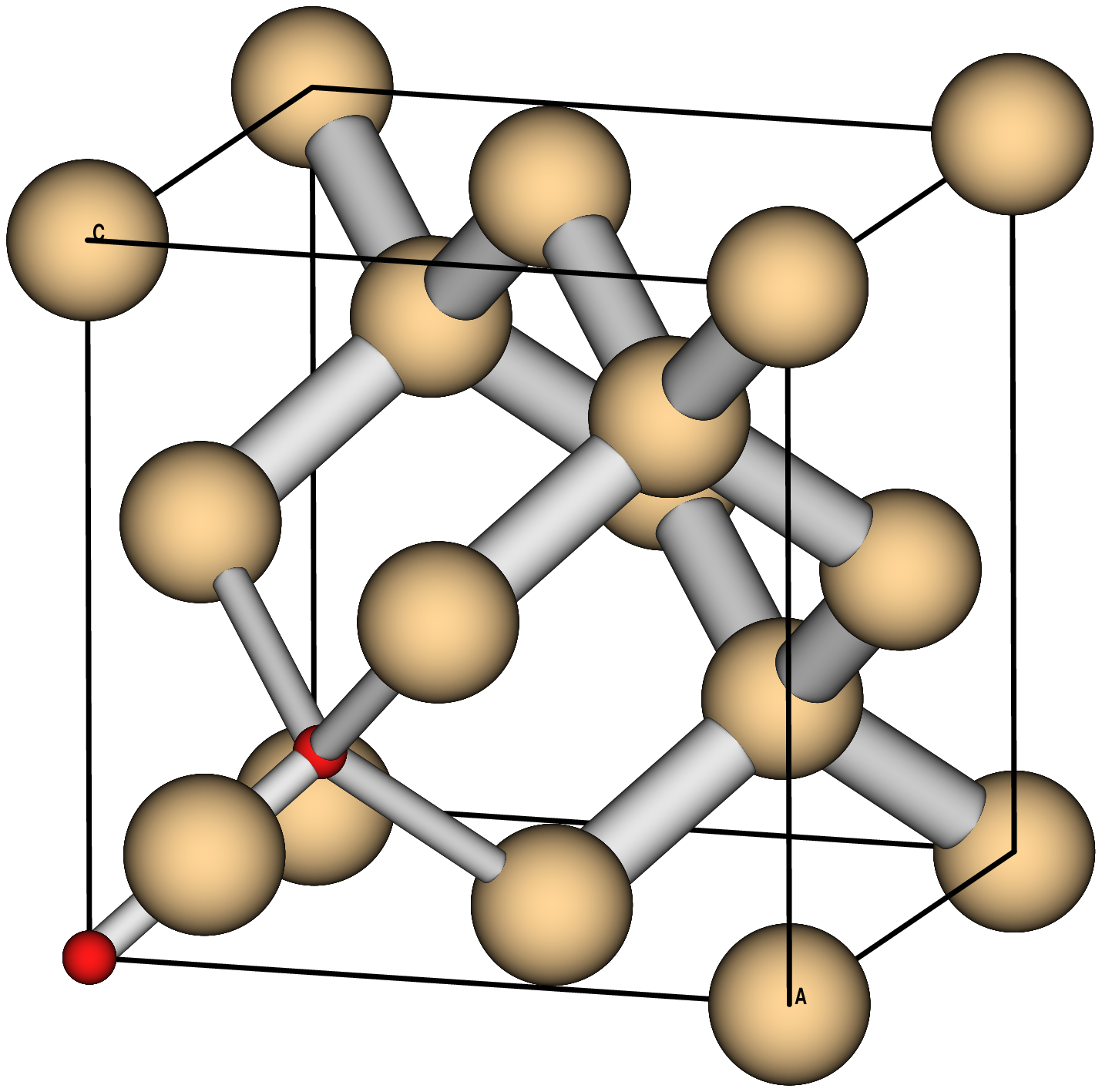}}
  \subfigure[$I_T$]{\label{fig:IT}\includegraphics[width=0.15\textwidth]{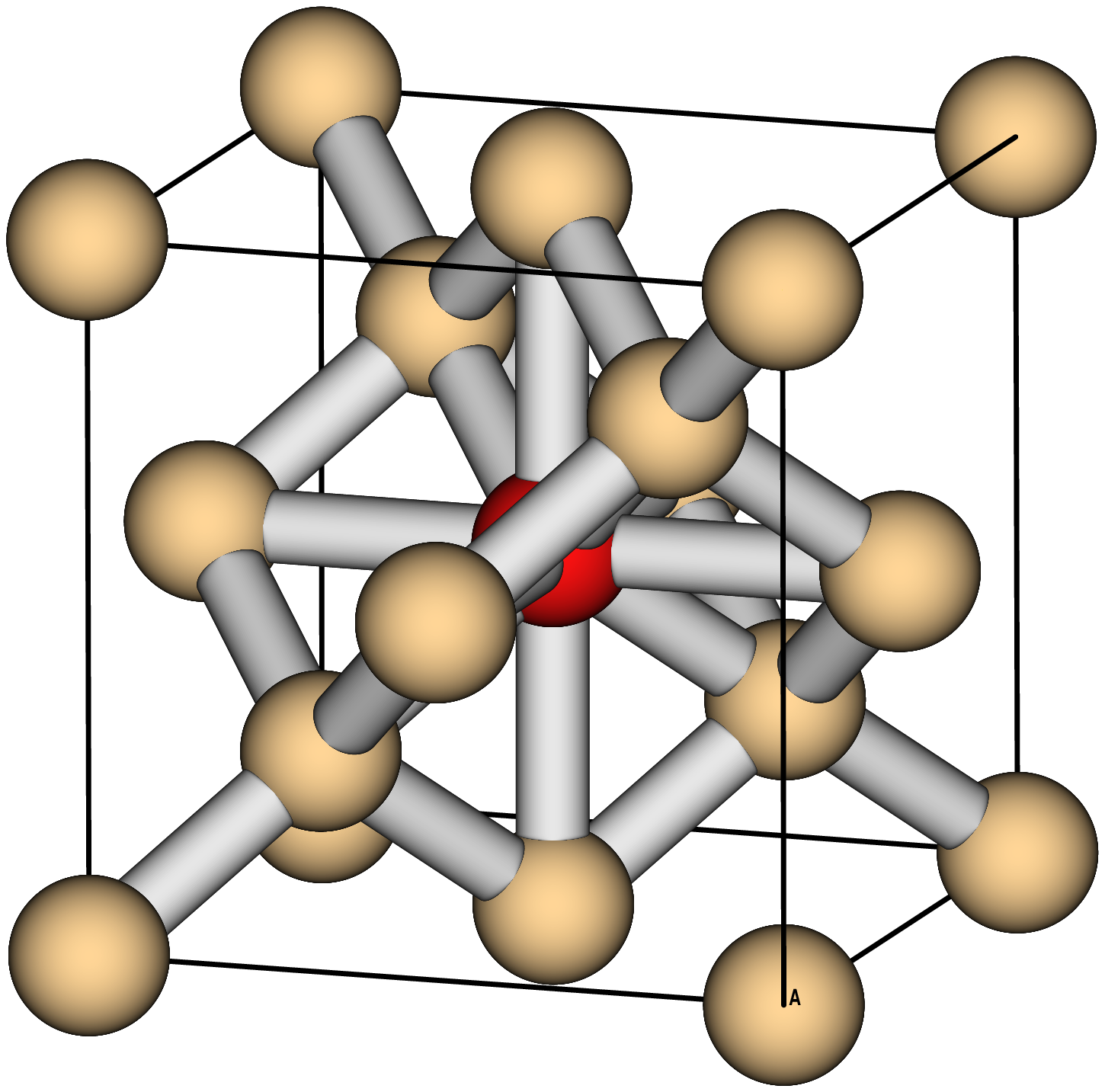}}
  \subfigure[$I_H$]{\label{fig:IH}\includegraphics[width=0.15\textwidth]{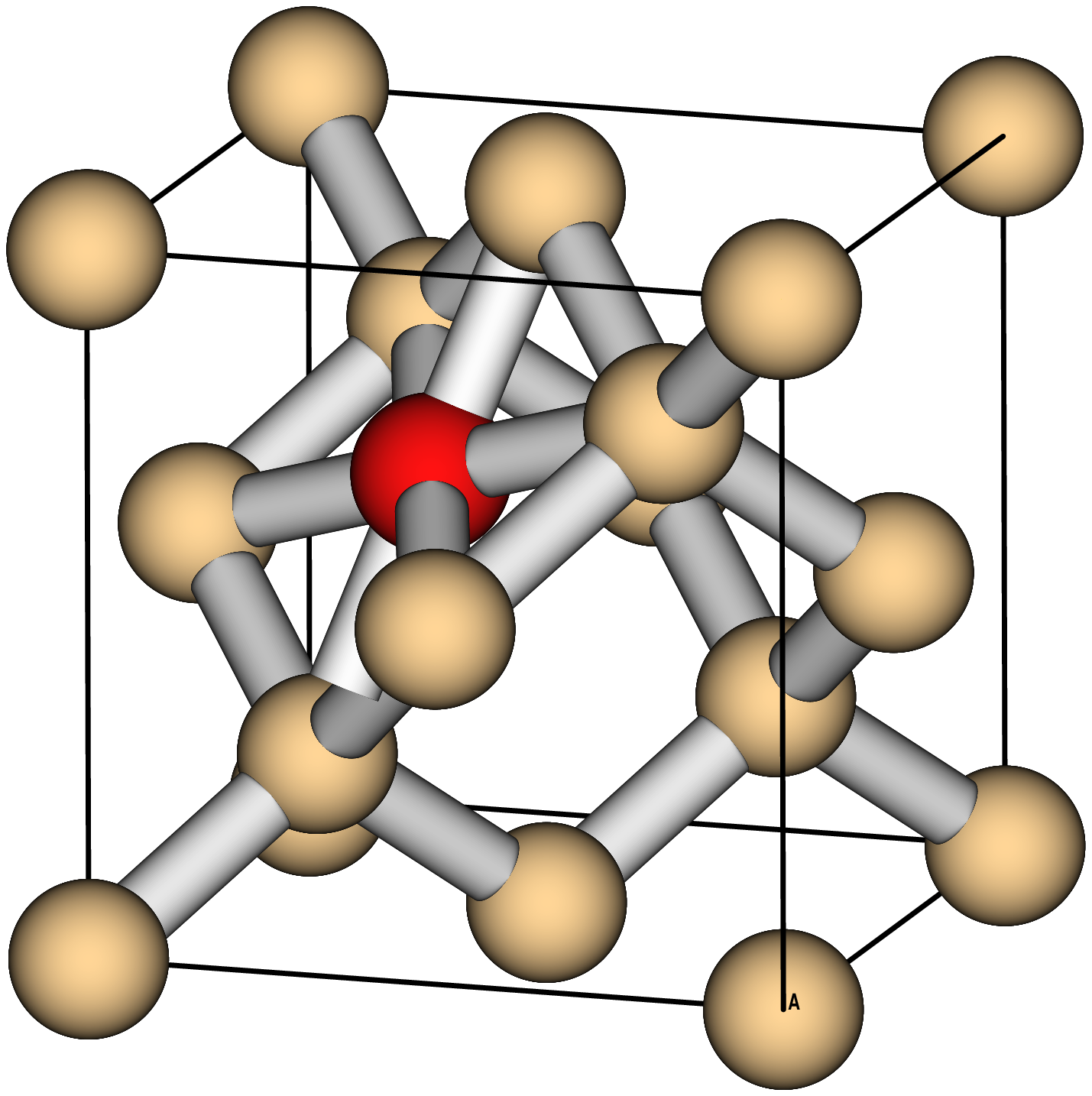}}\\
  \mycap{\label{fig:defectgeometries}Geometries of defects}{in
    relation to a conventional cubic unit cell of Si}
\end{figure}

Similar calculations were performed with supercells containing
defects.  In each case, the supercells were set at a particular volume
(scaled isotropically from the perfect lattice) and the ionic
coordinates were fully relaxed to build up a list of at least seven
energy--volume data points.  Then the same equation of state was fit
to the energy--volume data to find the enthalpy--pressure
relationship.  We define $h_\Si$ as the total enthalpy and $v_\Si$ as
the volume of a silicon atom in the perfect crystal. A supercell
containing 256 lattice sites and a vacancy includes 255 silicon atoms,
so we define the formation enthalpy of a vacancy as $h\formation_V
\equiv H_V[\Si_{255}] - 255 h_\Si$, and likewise the formation volume
is $v\formation_V \equiv V_V[\Si_{255}] - 255 \vol$.  Similarly, for
self-interstitials $h\formation_I \equiv H_I[\Si_{257}] - 257 h_\Si$
and $v\formation_I \equiv V_V[\Si_{257}] - 257 \vol$.  On the other
hand, the quantities of interest in elasticity are the relaxation
volumes $v\relaxation_X \equiv V_X[\Si_{256 \pm 1}] - 256 \vol$.  The
defects described in this work are shown in
Figure~\ref{fig:defectgeometries} and will be commented on later.

\section{Vacancies}
\label{sec:vacancies}
\begin{figure}[tbp]
  \centering \includegraphics[totalheight=0.25\textheight]{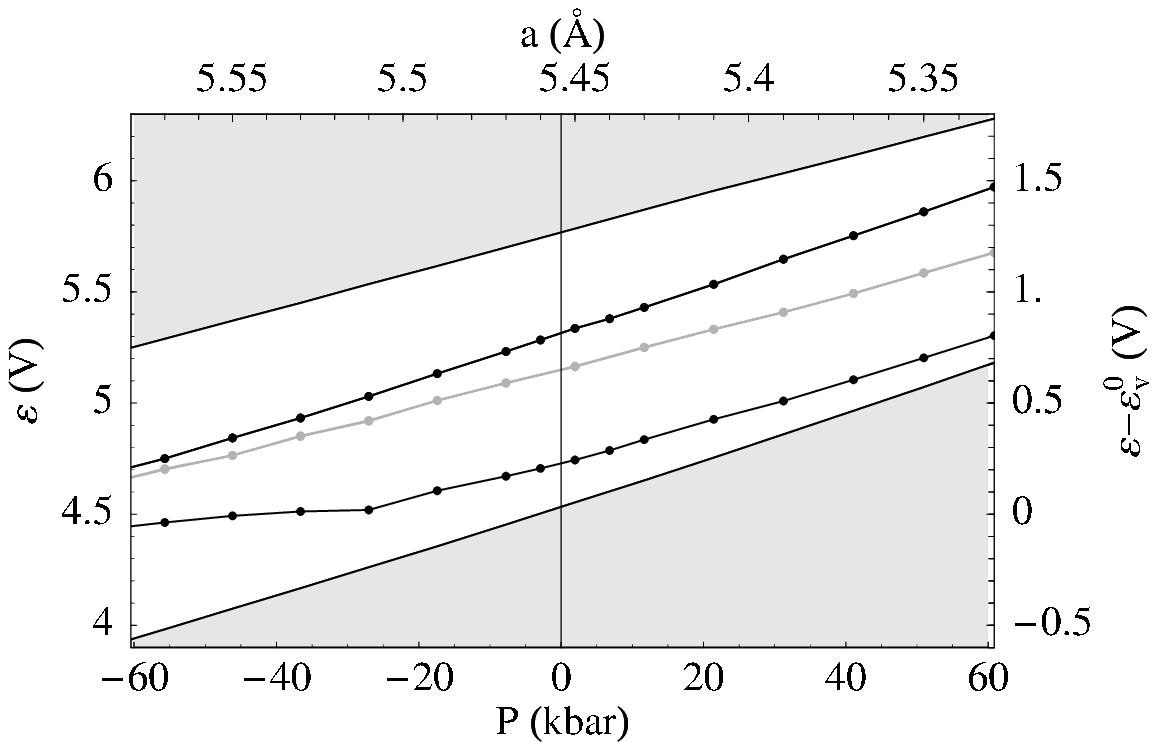}
  \mycap{\label{fig:JTeigs}Jahn--Teller splitting of energy
    eigenvalues.}{Shown are the band edges and the defect levels for
    $V_L^0$ with $T_d$ (\emph{gray}) and $V_L^0$ with $D_{2d}$
    (\emph{black}) symmetry.  Pressures are approximate.}
\end{figure}

Intuitively, the simplest point defect is a neutral lattice vacancy,
which we will label $V_L^0$: removing an atom from a perfect lattice.
However, the neighboring atoms will tend to rebond in ways that make
the defect less symmetric, particularly in a covalently bonded crystal
like silicon.  If the atoms are forced to maintain a $T_d$ symmetry,
the four neighbors draw in toward the center, pulling the rest of the
lattice with them.  However, the ground state involves a $T_d \to
D_{2d}$ symmetry-breaking relaxation explained in the early days of
quantum chemistry by Jahn and Teller \cite{Jahn37}. The Jahn--Teller
distortion of the neutral lattice vacancy in Si is now
well-established by experiment \cite{Watkins97} and theory
\cite{Tarnow93}.  We can see in Figure~\ref{fig:JTeigs} the splitting
of the  triply degenerate $T_2$ level (occupied by two electrons) into
a filled lower level and two degenerate empty upper levels, all still
in the band gap.

\begin{figure}[tbp]
  \centering
  \includegraphics{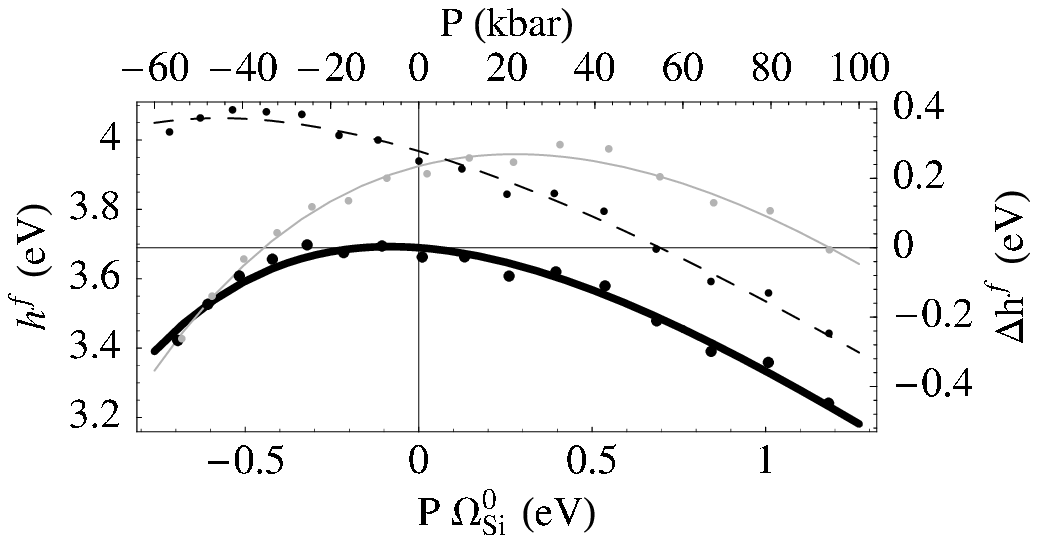}
  \mycap{\label{fig:hfV0}Effect of pressure on formation enthalpy
    of neutral vacancies.}{Shown here are curves for $V_L^0$ with
    both $T_d$ symmetry (\emph{gray}) and $D_{2d}$ (\emph{heavy}),
    as well as $V_B^0$ (\emph{dashed}).  Fermi level
    $\level\fermi$ fixed at intrinsic level $\level\intrinsic$.}
\end{figure}

Figure~\ref{fig:hfV0} shows that the Jahn--Teller distortion reduces
the enthalpy of the defect by about \unit{0.25}{\eV} at $P = 0$, and
results in a further contraction of $0.40 \vol$.  As a result,
the formation volume of a $T_d$ vacancy is positive in our
calculations, but that of a $D_{2d}$ vacancy is a small negative
value, $-0.07 \vol$ at $P = 0$, and vanishes at about
\unit{-6}{\kilo\bbar}.  This surprising result implies that
introducing vacancies into silicon (at $P = 0$) actually reduces the
volume of the system and increases its density.  A lattice with no
relaxation would have $v\formation_V = +\vol$, while perfect
relaxation (as with an incompressible liquid) would give
$v\formation_V = 0$.

\begin{figure}[tbp]
  \centering
  \centering \includegraphics[totalheight=0.25\textheight]{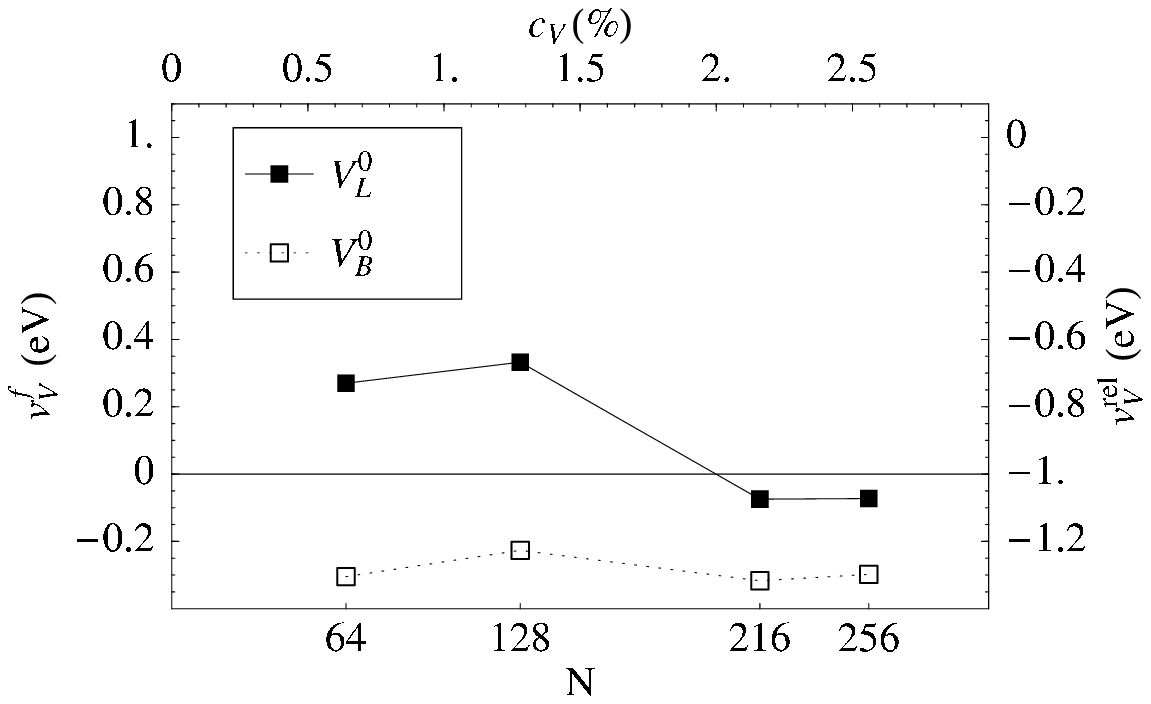}
  \mycap{\label{fig:VVsize}Effect of supercell size}{on neutral vacancy
    ($V^0$) formation volume at $P = 0$ and
    $\level\fermi = \level\intrinsic$.}
\end{figure}

The Jahn--Teller splitting vanishes under sufficiently large tensile
strain, but is approximately constant under compressive strain.  Other
first-principles calculations of $V_L^0$ (though performed only for $P
= 0$) have also reported small negative formation volumes
\cite{Antonelli98}.  In simulations reported in the literature, this
distortion is stable only when using supercells with more than 128
atoms \cite{Puska98}.  Our own calculations confirm this, with a
transition from tetrahedral to tetragonal geometry occurring between
128 and 216 atoms (Figure~\ref{fig:VVsize}).

Another high-symmetry configuration of $N-1$ atoms in a crystal with
$N$ sites is a so-called split vacancy, where one atom is at the bond
center between two empty sites, a configuration we label $V_B$.  This
can easily be seen as a transition point for vacancy migration.  Table
\ref{tab:vac} shows that $V_B$ has a fairly constant large negative
formation volume, about $-0.30 \vol$. However, in contrast to
the lattice vacancy, this is not due to a Jahn--Teller symmetry
breaking.  As a result, the formation volume of the six-coordinated
$V_B^0$ shows little change with supercell size
(Figure~\ref{fig:VVsize}) or pressure (Figure~\ref{fig:hfV0}).  From
these arguments it also follows that the vacancy migration enthalpy is
fairly constant for positive pressure, but increases for negative
pressures.

\begin{table}[tbp]
  \mycap{\label{tab:vac}Vacancy formation enthalpies and
  volumes}{at $P = 0$ and $\level\fermi =
  \level\intrinsic$.  $(v\formation_V  =  \vol + v\relaxation_V)$}
  \begin{ruledtabular}
  \begin{tabular}{r....}
 &
\multicolumn{2}{c}{$V_L^q$}&
\multicolumn{2}{c}{$V_B^q$}\\
$q$ &
\ccol{$h\formation (\unit{\eV})$} &
\ccol{$v\formation (\vol)$} &
\ccol{$h\formation (\unit{\eV})$} &
\ccol{$v\formation (\vol)$}
\\[2mm]
\hline\\[-3mm]
$-2$&  4.33&  0.08&  4.14& -0.27\\
$-1$&  3.87&  0.01&  3.89& -0.30\\
$ 0$&  \textbf{3}.\textbf{69}& \textbf{--0}.\textbf{07}&  3.97& -0.30\\
$+1$&  4.07&  0.21&  4.29& -0.37\\
$+2$&  4.55&  0.43&  4.90& -0.42\\
\end{tabular}
  \end{ruledtabular}
\end{table}

\begin{figure*}[tbp]
  \centering
  \subfigure[Vacancies]{\label{fig:FermiV}\includegraphics{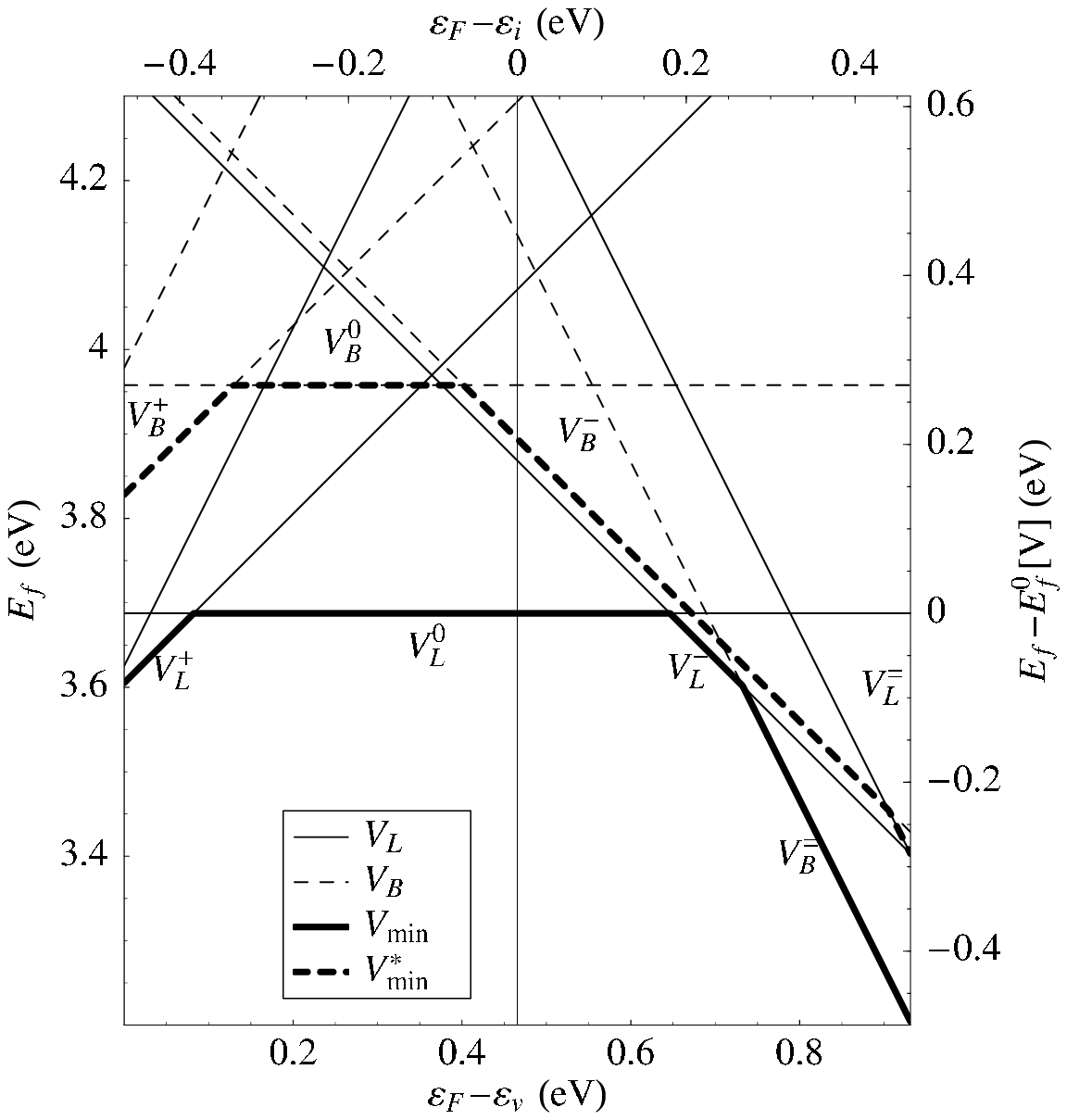}}
  \subfigure[Interstitials]{\label{fig:FermiI}\includegraphics{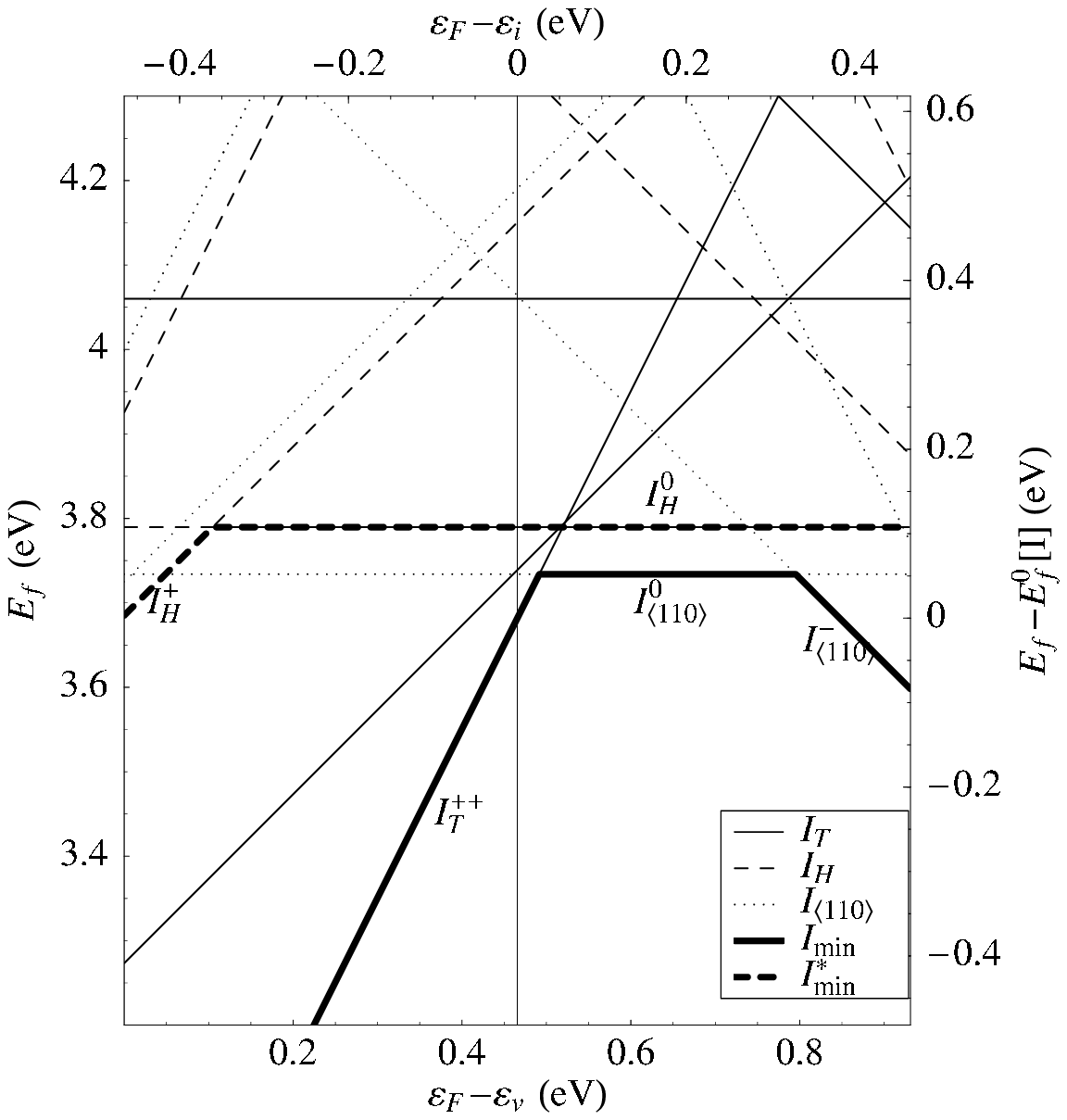}}
  \mycap{\label{fig:Fermi}Intrinsic defect formation enthalpies \textit{vs.}  Fermi
    level}{at $P = 0$.  The heavy solid line indicates the lowest
    formation enthalpy $h\formation$ for a defect at that Fermi
    level.  The heavy dashed line indicates the second-lowest
    enthalpy, which is an estimate of $h\diffusion$, the
    activation enthalpy of the transport capacity.  The difference
    between the lines would then be the activation enthalpy of
    migration, $h\migration$.  Note that slope is proportional to
    the charge of the defect.  A vertical line marks the
    calculated location of the intrinsic Fermi level
    $\level\intrinsic$.}
\end{figure*}

We have also calculated the formation enthalpies and volumes of
charged vacancies (Table~\ref{tab:vac}), since charged defects are
expected to play an important role in doped silicon.
Figure~\ref{fig:FermiV} shows that the neutral lattice-centered
vacancy $V_L^0$ is the stable configuration over the broadest range of
Fermi levels within the band gap (including the intrinsic level
$\level\intrinsic = (\level\valence + \level\conduction)/2$) and has a
formation enthalpy of \unit{3.69}{\eV} at $P = 0$.  The bond-centered
split vacancy $V_B^0$ is \unit{0.27}{\eV} higher.

As mentioned above, $V_B$ is the transition point for a vacancy to
migrate from one $V_L$ configuration to the next. However, our
calculations reveal a surprising twist: The negative split vacancy
$V_B^-$ has almost the same energy as the negative lattice vacancy
$V_L^-$, and at high Fermi levels, the ground state is in fact the
split vacancy (the doubly negative split vacancy is even lower in
energy than the lattice vacancy). This phenomenon was predicted
decades ago by \citet{Bourgoin75}, but is not well-known by all who
work with silicon diffusion, since most industrially useful processes
involve elevated temperatures.  This reversal may be due to the
greater number of bonds that can accept extra electrons in the case of
the split vacancy.  Since the displaced atom in the split vacancy has
four valence electrons and six neighbors, giving the system two extra
electrons allows this atom to form six equally strained bonds with all
of its neighbors.  In other words, the minimum-enthalpy geometry in
one charge state ($-2$) is a saddle point in a different charge state
($0$), and vice versa.  This crossover in potential energy surface is
the requirement for the Bourgoin mechanism of athermal (electronically
or optically activated) diffusion to take place
\cite{Bourgoin72,Bourgoin75}.  Experiments have demonstrated that
vacancies in $n$-type silicon can diffuse at room temperature or even
cryogenic temperature when subjected to optical or electronic
excitation \cite{Watkins97,Larsen99}.


\section{Self-Interstitials}
\label{sec:self-interstitials}
We have calculated the enthalpies of the three interstitial geometries
found to have the lowest enthalpies in previous work: tetrahedral
($I_T$), hexagonal ($I_H$), and split-$\braket{110}$
($I_{\braket{110}}$).  Our self-interstitial calculations are
summarized in Table~\ref{tab:int}.  In contrast to the strong pressure
and Fermi level sensitivity of vacancies, self-interstitials present a
somewhat simpler picture.  The lowest enthalpy interstitial at the
intrinsic Fermi level and $P = 0$ is $I_T^{++}$, with a formation
enthalpy of \unit{3.68}{\eV} and a formation volume of $-0.51 \vol$.
(See the Methods section for an explanation of how formation
enthalpies and volumes are defined in our calculations.)

We can explain the stability region of the various self-interstitial
configurations, by plotting their formation enthalpies as a function
of Fermi level (Figure~\ref{fig:FermiI}). We see that the $I_T^{++}$
configuration is stable mainly in $p$-type Si, while at slightly
elevated Fermi levels ($n$ doping) the $I_{\braket{110}}^0$ split
interstitial is stabilized. A strong negative-$U$ effect is seen,
where the interstitial is nowhere stable in the $+1$ charge state.


\begin{table}[tbp]
  \mycap{\label{tab:int}Self-interstitial formation enthalpies and
  volumes}{at $P = 0$ and $\level\fermi =
  \level\intrinsic$.  ($v\formation_I  = -\vol + v\relaxation_I$)}
  \begin{ruledtabular}
  \begin{tabular}{r......}
 &
\multicolumn{2}{c}{$I_T^q$}&
\multicolumn{2}{c}{$I_H^q$}&
\multicolumn{2}{c}{$I_{\braket{110}}^q$}\\
$q$ &
 \ccol{$h\formation$ (\unit{\eV})} &
 \ccol{$v\formation (\vol)$} &
 \ccol{$h\formation$ (\unit{\eV})} &
 \ccol{$v\formation (\vol)$} &
 \ccol{$h\formation$ (\unit{\eV})} &
 \ccol{$v\formation (\vol)$}
\\[2mm]
\hline\\[-3mm]
$-2$&  5.37& -0.73&  5.12& -0.49&  4.70& -0.45\\
$-1$&  4.61& -0.67&  4.34& -0.43&  4.06& -0.42\\
$ 0$&  4.06& -0.63&  3.79& -0.38&  3.73& -0.41\\
$+1$&  3.73& -0.60&  4.15& -0.41&  4.19& -0.45\\
$+2$&  \textbf{3}.\textbf{68}& \textbf{--0}.\textbf{51}&  4.85& -0.49&  4.92& -0.51\\
\end{tabular}
  \end{ruledtabular}
\end{table}


Much like the situation with vacancies, high-symmetry interstitial
geometries can be saddle points for interstitial migration.  In
particular, the hexagonal interstitial is usually assumed to be the
migration point for interstitial diffusion.  Figure~\ref{fig:FermiI}
suggests that this is true for Si at almost any doping; hence
interstitial migration in $n$-doped Si ($I_{\braket{110}}^0
\rightarrow I_H^0 \rightarrow I_{\braket{110}}^0$) has a migration
barrier of only \unit{0.06}{\eV}, while in $p$-doped Si the migration
mechanism $I_T^{++} \rightarrow I_H^+ \rightarrow I_T^{++}$ leads to a
barrier that increases linearly with the doping level, thus causing a
drastic slow-down of interstitial migration. 

A particularly interesting regime is that of the Fermi level being
close to the midgap, where the $I_T^{++}$ is only slightly lower in
energy than the neutral $I_{\braket{110}}^0$. It is easy to see that
under such conditions, addition of electrons \eg through electron
irradiation, can lead to a mechanism without any activation barrier.
This so-called Bourgoin mechanism of electrically, rather than
thermally, activated diffusion is observed experimentally at cryogenic
temperatures during electron irradiation of intrinsic Si.  One of the
implications of this is that annealing after ion implantation should
be significantly faster in $n$-type silicon than in $p$-type.

\section{Frenkel Pairs}
\label{sec:frenkel-pairs}
Combining the formation volumes of the most stable vacancy and
self-interstitial configurations indicates that a Frenkel pair ($V +
I$) should have a formation enthalpy of \unit{7.39}{\eV} and a
formation volume of $-0.68 \vol$.  At elevated Fermi levels (\ie{}
high $n$ doping), the stable self-interstitial species becomes
$I_{\braket{110}}^0$, which would lead to a Frenkel pair formation
enthalpy of \unit{7.44}{\eV} and formation volume of $-0.48 \vol$.

On the other hand, Huang diffuse x-ray scattering experiments
performed by Ehrhart \etal\ \cite{Ehrhart97,Partyka01} have been taken
to conclude that the formation volumes of Frenkel pairs (produced by
\unit{2.5}{\mega\eV} $\electron$ or \unit{4.5}{\kilo\eV} He
implantation) are quite small, $0.1 \vol$ or less.  This result
is based on the assumption that at cryogenic temperatures, electron
irradiation generates a substantial number of Frenkel pairs with a
separation distance of only about \unit{8}{\Ao}.

The disagreement of the Erhart experiments with our first-principles
calculations suggests that Frenkel pairs with short separation
distances behave differently from the isolated defects.  This is not
an unreasonable theory when considering our previous findings (see
section Vacancies) of the sensitivity of the formation volume of the
vacancy to its concentration (see Fig. 4).  To see whether vacancies
and interstitials at close proximity can behave anomalously, we
constructed supercells containing a self-interstitial and a vacancy
approximately \unit{8}{\Ao} apart.  Two different interstitial
geometries were used, $I_T$ and $I_{\braket{110}}$.  Only neutral
supercells were considered, since the claim of Ehrhart \etal{} is that
irradiation of silicon creates large concentrations of Frenkel pairs
that are either entirely neutral or form donor--acceptor pairs.  In
this case, the Frenkel pair would not consist of $V_L^0 + I_T^{++}$,
but perhaps of $V_L^0 + I_{\braket{110}}^{0}$, $V_L^- + I_T^+$, or
$V_L^{--} + I_T^{++}$.  Formation volumes are calculated as before.
The results are shown in Table~\ref{tab:frenkel}, together with summed
values for the isolated defects.

\begin{table}[htbp]
  \mycap{\label{tab:frenkel} Frenkel pair enthalpies and volumes of
    formation.}{Values for isolated defects are summed for
    convenience.}
  \begin{ruledtabular}
    \begin{tabular}{cc..}
      geometry &
      $r (\Ao)$ &
      \ccol{$h\formation (\eV)$} &
      \ccol{$v\formation (\vol)$} \\[2mm]
      \hline\\[-3mm]
      $(V_B I_{T})^0$&                  5.88&  6.88&  -0.89\\[1mm]
      $(V_L I_{T})^0$&                  8.83&  7.18&  -0.63\\[1mm]
      $(V_L I_{\braket{110}})^0$&       8.21&  7.31&  -0.48\\[1mm]
      \hline\\[-3mm]
      $V_L^0 + I_{\braket{110}}^0$& $\infty$&  7.44&  -0.48\\[1mm]
      $V_L^{0} + I_{T}^{++}$&       $\infty$&  7.39&  -0.68\\[1mm]
      $V_L^{--} + I_{T}^{++}$&      $\infty$&  8.04&  -0.62\\[1mm]
    \end{tabular}
  \end{ruledtabular}
\end{table}

The formation volume for $V_L^0 + I_{\braket{110}}^0$ is nearly the
same whether both defects are in the same supercell or calculated
separately.  The formation enthalpy is a bit lower for the pair than
for the isolated defects, which indicates the strength of the their
attraction.  Interestingly, in the case of $(V_L I_{T})^0$, the
relaxation of the atoms surrounding the vacancy took the form of a
bent square, rather than two pairs seen in the isolated vacancy, even
when the symmetry was broken in the direction of the expected
geometry.  In summary, for either the $T$ or $\braket{110}$ geometry,
the Frenkel pair formation volume is a sizable negative number,
contrary to the experimental results of Erhart et al.

In light of this, the assertion of Erhart et al. that irradiation of
Si at cryogenic temperatures will create a large number of Frenkel
pairs with short separation distances is inconsistent with
first-principles calculations.  Instead, our calculations suggest that
far more defect clustering occurs than would be expected from purely
classical migration mechanism of neutral defects.  For samples
irradiated by electrons or light, the Bourgoin mechanism described in
previously can lead to athermal diffusion.  On the other hand,
radiation damage from ions can deposit enough energy in a small volume
to directly create clusters of defects or even amorphous pockets and
local melting despite cryogenic background temperatures.

Besides the possibility of clustering even at low temperatures, $V-V$
interactions may provide another possible explanation of the low
observed Frenkel pair formation volume.  The largest supercell in our
calculations that stabilized the $T_d$ symmetry was the 128-site
supercell, having $h\formation[V_L^0] = \unit{3.51}{\eV}$
 and $v\formation[V_L^0] = 0.34 \vol$,
(Figure~\ref{fig:VVsize}), with larger values of each expected with a
larger supercell size.  This would imply that $h\formation[V + I] =
\unit{7.21}{\eV}$ and $v\formation[V + I] = -0.24 \vol$.
Interestingly, summing with $I_{\braket{110}}^0$ instead of $I_T^{++}$
would lead to $h\formation[V + I] = \unit{7.26}{\eV}$ and
$v\formation[V + I] = -0.04 \vol$.  A single defect in a 100-site
supercell corresponds to a defect concentration of 1\%.  At
sufficiently high vacancy concentrations, $V-V$ elastic interactions
may destabilize the Jahn--Teller distortion, leading to a positive
vacancy formation volume and a small Frenkel pair formation volume, as
well as a lower formation enthalpy.  The critical concentration to
destabilize the Jahn--Teller distortion for actual vacancies in
silicon need not lie in the interval $\frac{1}{128} < c_V <
\frac{1}{216}$.

Some earlier experiments claimed to find that $c_V D_V = c_I D_I$
at $T = \unit{800}{\celsius}$ \cite{Tan85} or $T =
\unit{1000}{\celsius}$ \cite{Bracht98}.  Near that temperature, the
effective activation enthalpy of Si self-diffusion should be close to
$\half(h\diffusion[V] + h\diffusion[I])$.  Our calculations indicate
$h\formation[V] = \unit{3.69}{\eV}$ and $h\formation[I] =
\unit{3.68}{\eV}$ under conditions of no doping or applied pressure.
Our estimate of $h\migration[V] = \unit{0.27}{\eV}$ appears
reasonable, but we take $h\migration[I] = \unit{0.22}{\eV}$ based on a
study of low-symmetry pathways not investigated here \cite{Needs99}.
Thus we arrive at $h\diffusion[V] = \unit{3.96}{\eV}$ and
$h\diffusion[I] = \unit{3.90}{\eV}$.  These estimates are rather lower
than most experimental reports.  DFT methods predict migration
enthalpies more accurately than they do formation enthalpies.
Probably the most accurate method applied to the calculation of
intrinsic defects in Si is diffusion quantum Monte Carlo (DMC), which
gives $h\formation[I_H^0] = \unit{4.82}{\eV}$ \cite{Leung99},
\unit{1.02}{\eV} greater than the PW91 value, and leads to
$h\diffusion[I] = \unit{5.04}{\eV}$.

\section{Pressure--Fermi level stability diagrams}
\label{sec:pressure-fermi-level}
Examining the combined effects of changing pressure and Fermi
level requires finding the enthalpies and volumes of formation of
electrons and holes in Si as a function of pressure.  The
enthalpies are essentially the conduction band minimum
\begin{equation}
 h\formation\electrons(P) \equiv H[\Si^{-1}_N](P) - N h_\Si(P)
\end{equation}
and the valence band maximum
\begin{equation}
 h\formation\holes(P) \equiv N h_\Si(P) - H[\Si^{+1}_N](P).
\end{equation}
and the volumes are
\begin{equation}
  v\formation\electrons(P) \equiv V[\Si^{-1}_N](P) - N \vol(P)
  = \deriv{\level\conduction(P)}{P}
\end{equation}
and
\begin{equation}
  -v\formation\holes(P) \equiv N \vol(P) - V[\Si^{+1}_N](P)
  = \deriv{\level\valence(P)}{P}.
\end{equation}
At $P = 0$, we have $v\formation\electrons = 0.679 \vol$ and
$v\formation\holes = -0.788 \vol$, a sizable effect.

\begin{figure}[tbp]
  \centering
  \subfigure[Vacancy]{\label{fig:Vfsurf}\includegraphics[width=0.49\columnwidth]{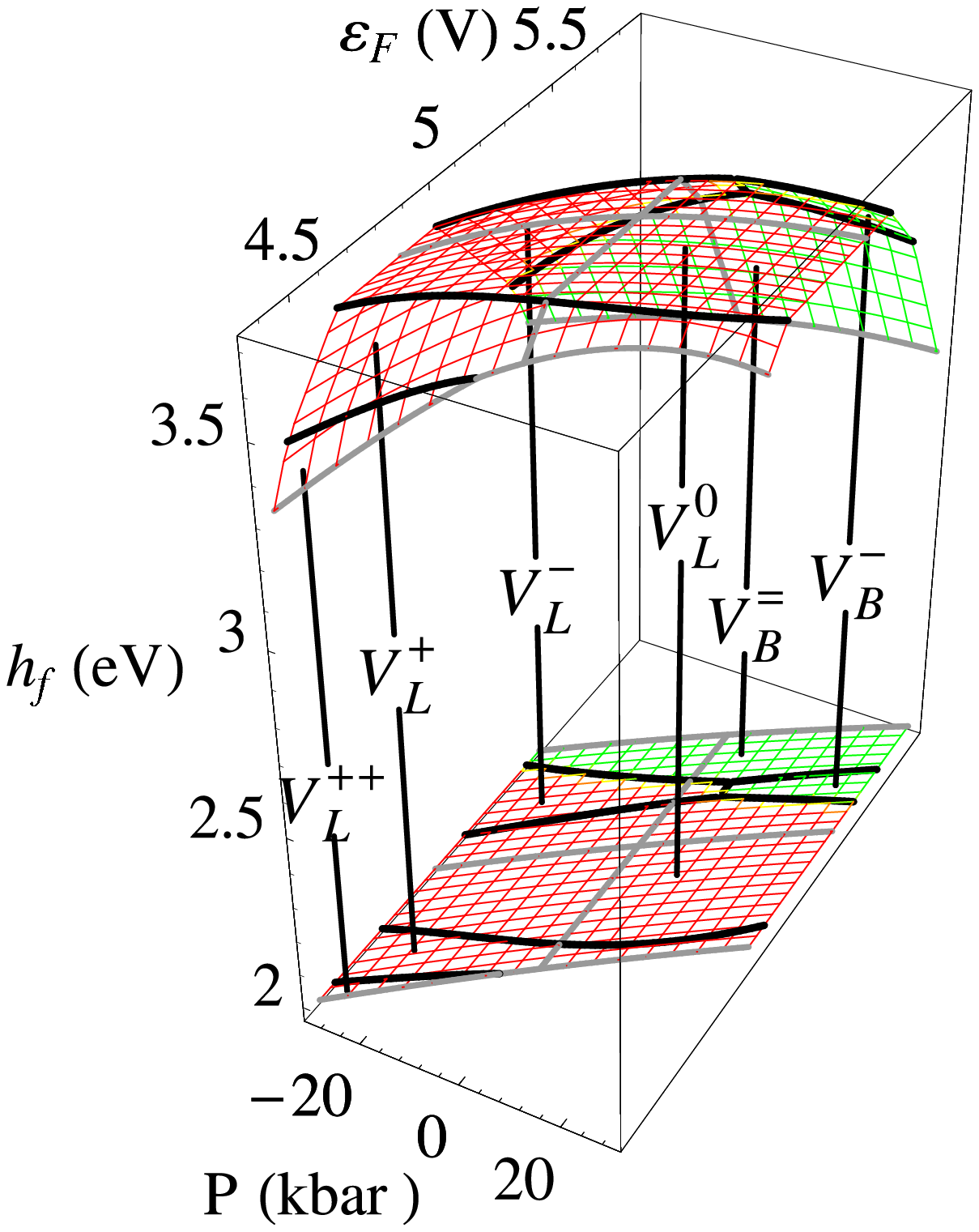}}
  \subfigure[Interstitial]{\label{fig:Ifsurf}\includegraphics[width=0.49\columnwidth]{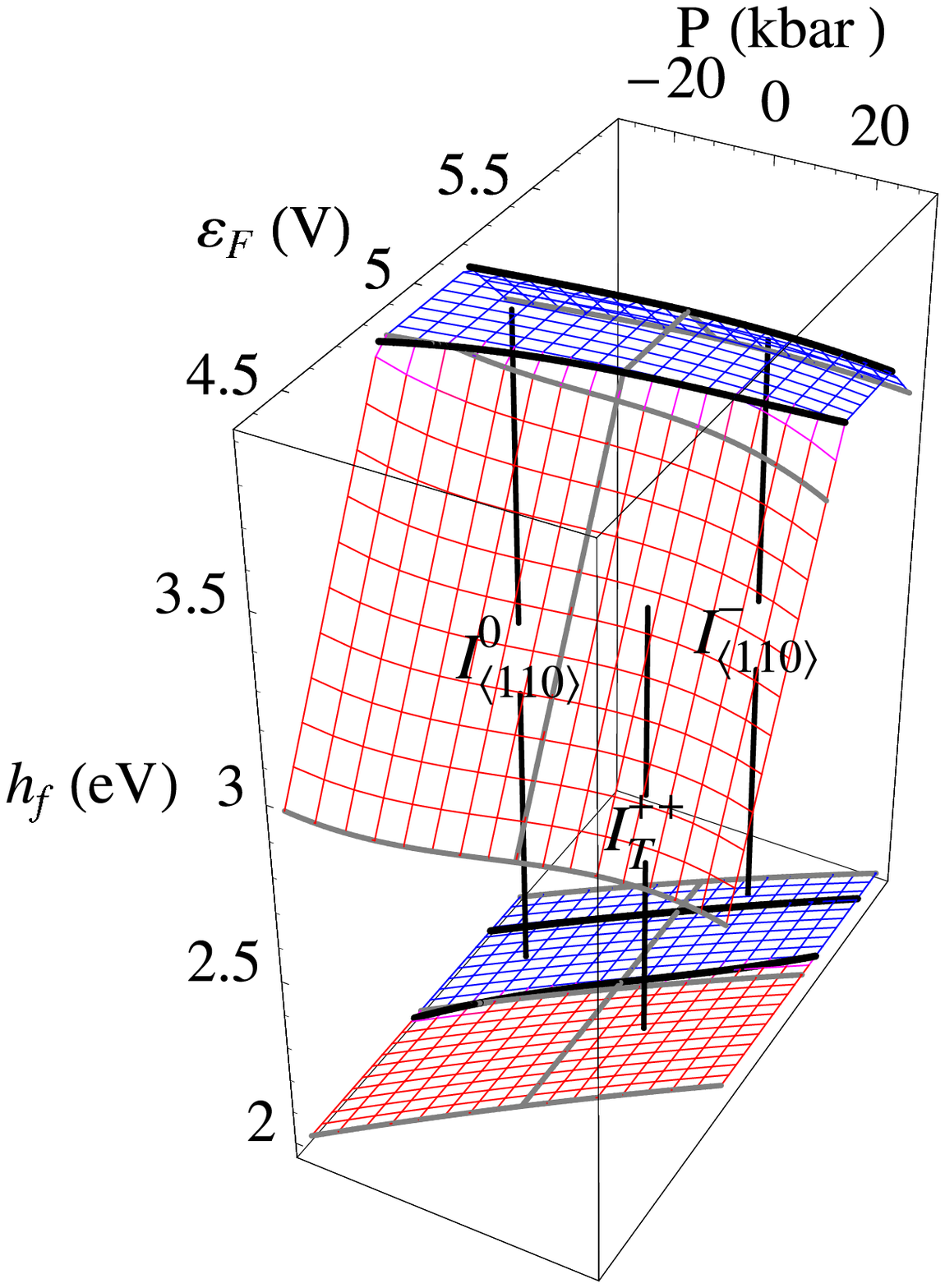}}\\
  \mycap{\label{fig:minsurfs}Minimum formation enthalpy of defects
    as a function of $P$ and $\level\fermi)$.}{The intrinsic Fermi
    level is indicated as a gray line.}
\end{figure}
\begin{figure}[tbp]
  \centering
  \subfigure[Vacancy]{\label{fig:Vfphase}\includegraphics[width=0.49\columnwidth]{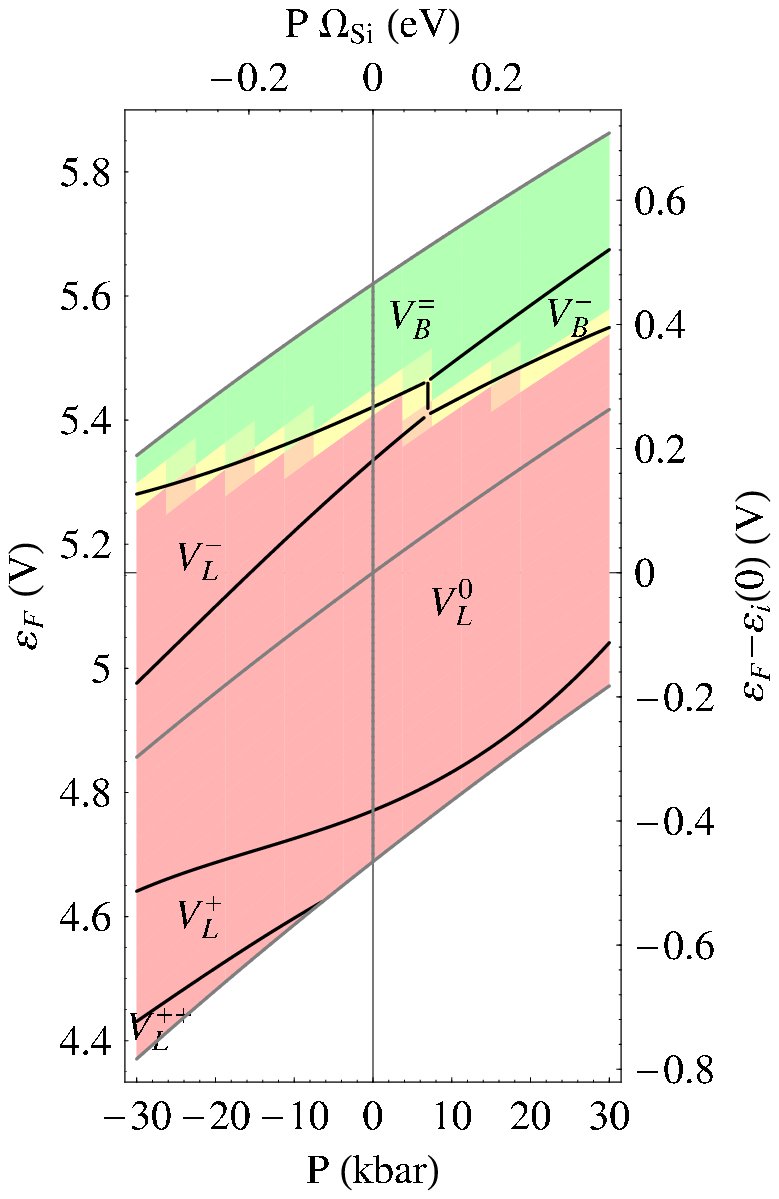}}
  \subfigure[Interstitial]{\label{fig:Ifphase}\includegraphics[width=0.49\columnwidth]{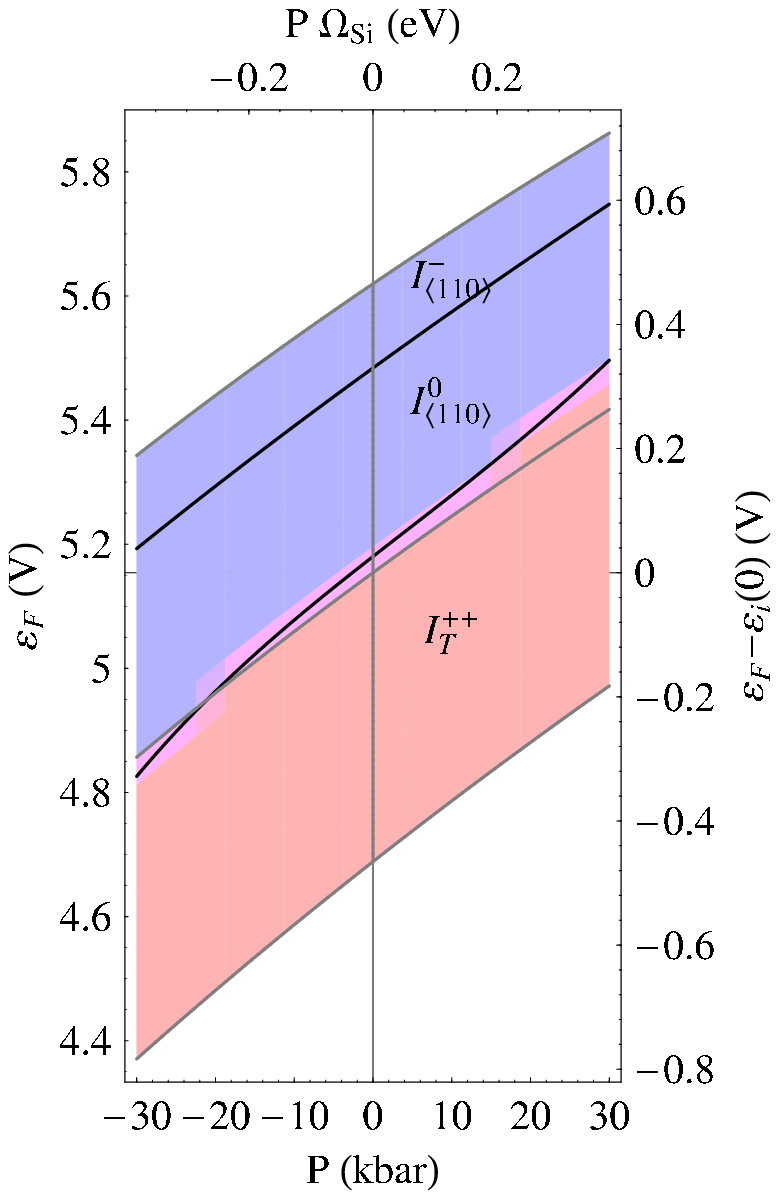}}\\
  \mycap{\label{fig:phases}$P-\level\fermi$ stability diagram of
    intrinsic defects in Si}{formed by projecting the surface of
    minimum enthalpy in Figure~\ref{fig:minsurfs} for $V_L$ (red),
    $V_B$ (green), $I_T$ (red), $I_H$ (green), and $I_{\braket{110}}$
    (blue). The intrinsic Fermi level is indicated as a gray line.}
\end{figure}

Since we have the pressure dependence of all these quantities, we can
essentially sweep Figure~\ref{fig:Fermi} over a finite pressure range,
where the domain of the plot is the band gap.  Showing all the
enthalpy surfaces
obscures the most important information, the lowest enthalpy defects.
More useful is a plot of just the minimum enthalpy required to form a
defect at a particular pressure and Fermi level
(Figure~\ref{fig:minsurfs}).  The vacancy surface
(Figure~\ref{fig:Vfsurf}) shows clear changes with pressure and marked
curvature, indicating a variable formation volume, while the shape of
the interstitial surface (Figure~\ref{fig:Ifsurf}) is close to being a
prism.  The location of the Fermi level is important for
interstitials, but the pressure is not.  Projecting these surfaces
down onto the $(P,\level\fermi)$ plane reveals other differences.  The
vacancy stability diagram (Figure~\ref{fig:Vfphase}) displays a wealth
of features: the instability of $V_L^{++}$ under pressure, the
transition from $V_L^-$ to $V_B^-$, and in general the large changes
in vacancy levels caused by pressure.  None of these features are
evident for interstitials (Figure~\ref{fig:Ifphase}).


\section{Relaxation volume tensors}
\label{sec:relax-volume-tens}
The stress state in integrated circuits is seldom hydrostatic.  On the
scale of the wafer it is typically biaxial due to surface oxidation or
nitridation, but at a sub-micron scale becomes quite complicated, with
enormous stress gradients.  Aziz \cite{Aziz01} has described the
effect of the formation volume of defects in a film under biaxial
stress, but points out that a full treatment requires more information
about a defect than just its scalar volume.  Many calculations of
defects in silicon point out changes in the positions of the
neighboring atoms, but this can provide only qualitative, not
quantitative information about the complete elastic distortion caused
by the defect.  The present work fills this gap.

Among the features of the code used here is full relaxation of the
vectors defining the supercell.  This tells us how the shape as
well as the size of the supercell change when a defect is formed.
Were we dealing with a simple cubic cell changing isotropically
from edge length $L_0$ to $L$, we could define the relaxation
volume as
\begin{equation}
  \label{eq:relaxscal}
  v\relaxation
  = L_0^3 \ln \left(\frac{L}{L_0}\right)^3
  \approx L_0^3 \left(\frac{L - L_0}{L_0}\right)^3 .
\end{equation}
Instead, we have three non-orthogonal vectors (forming a
transformation matrix $\mathmat{L} = L_{ij}$) changing in
arbitrary directions.  We may generalize \eqref{eq:relaxscal} as
\begin{equation}
  \label{eq:relaxtens}
  \mathmat{v}\relaxation
  = \det(\mathmat{L}_0) \ln \left(\mathmat{L}_0^{-1} \mathmat{L} \right)
  \approx \det(\mathmat{L}_0) \mathmat{L}_0^{-1} \left(\mathmat{L} - \mathmat{L}_0^{-1}\right)
\end{equation}

\begin{table}[tbp]
  \mycap{\label{tab:relaxtens}Relaxation volume tensors
    $v\relaxation_{ij}$ of neutral defects in Si}{at $\sigma_{ij} = 0$
    and $\level\fermi = \level\intrinsic$.  The scalar part is
    indicated as well as the eigenvalues and corresponding principal
    directions.  Scalar volumes in parentheses reflect the
    relaxation volumes calculated using the boundary
    conditions described previously (\ie{} purely dilatational
    strain of the supercell).}
\begin{ruledtabular}
\begin{tabular}{c.c|c.c}
vacancy &
\ccol{$v_i (\vol)$} &
\ccol{$\mathvec{n}_i$} &
interstitial &
\ccol{$v_i (\vol)$} &
\ccol{$\mathvec{n}_i$} \\[1mm]
\hline\\[-3mm]
  $V_L^0$& -0.83& $[110]$& $I_{\braket{110}}^0$& +0.57& $[110]$\\
  $-1.02$& -0.82& $[1\overline{1}0]$&   $+0.68$& +0.09& $[001]$\\
$(-1.07)$& +0.63& $[001]$&            $(+0.59)$& +0.02& $[1\overline{1}0]$\\
&&&&&\\[-1mm]
  $V_B^0$& -0.95& $[111]$&              $I_H^0$& +0.23& $[2\overline{11}]$\\
  $-1.30$& -0.18& $[2\overline{11}]$&   $+0.64$& +0.23& $[01\overline{1}]$\\
$(-1.30)$& -0.18& $[01\overline{1}]$& $(+0.62)$& +0.19& $[111]$\\
&&&&&\\[-1mm]
            &      &        &            $I_T^0$        & +0.12& $[111]$\\
            &      &        &           $+0.37$& +0.12& $[2\overline{11}]$\\
            &      &        &         $(+0.37)$& +0.12& $[01\overline{1}]$
\end{tabular}
\end{ruledtabular}
\end{table}

The relaxation volume tensors defined in this way are equivalent
to the strain dipole tensors
\begin{equation}
  \lambda_{ij} = v\relaxation_{ij} / \vol
\end{equation}
discussed by Nowick and Berry \cite{Nowick72} and closely related
to the piezo-spectroscopic elastic dipole tensors
\begin{equation}
  P_{ij} = C_{ijkl} v\relaxation_{kl}
\end{equation}
of Kr{\"o}ner \cite{Kroener58} (where $C_{ijkl}$ is the elastic
modulus tensor of the material).

\begin{figure*}[tbp]
  \centering
  \subfigure[$V_L^0$]{\label{fig:VL0vrelxxyyzz}
    \includegraphics[width=0.19\textwidth]{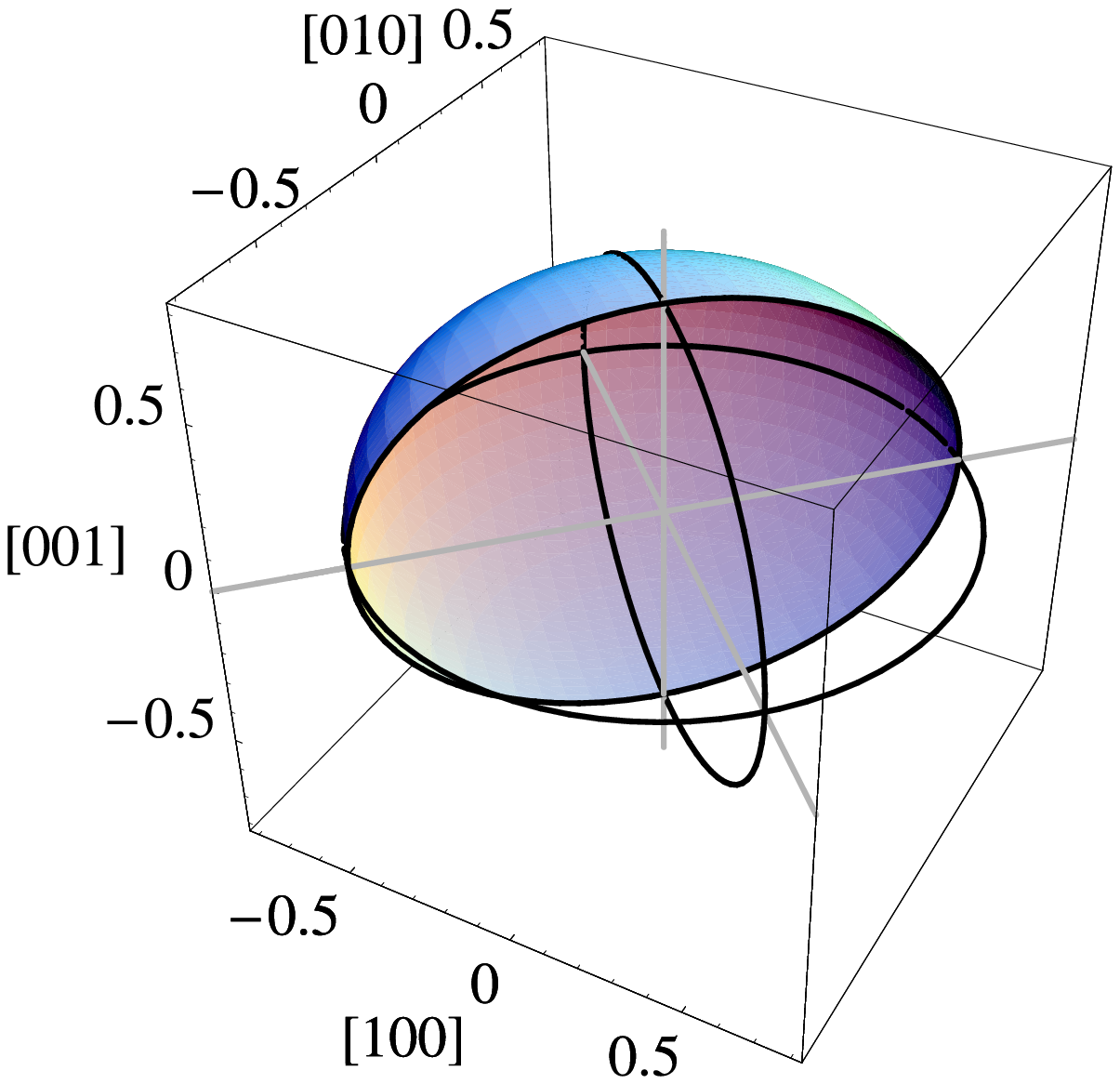}}
  \subfigure[$I_{\braket{110}}^0$]{\label{fig:IX0vrelxxyyzz}
    \includegraphics[width=0.19\textwidth]{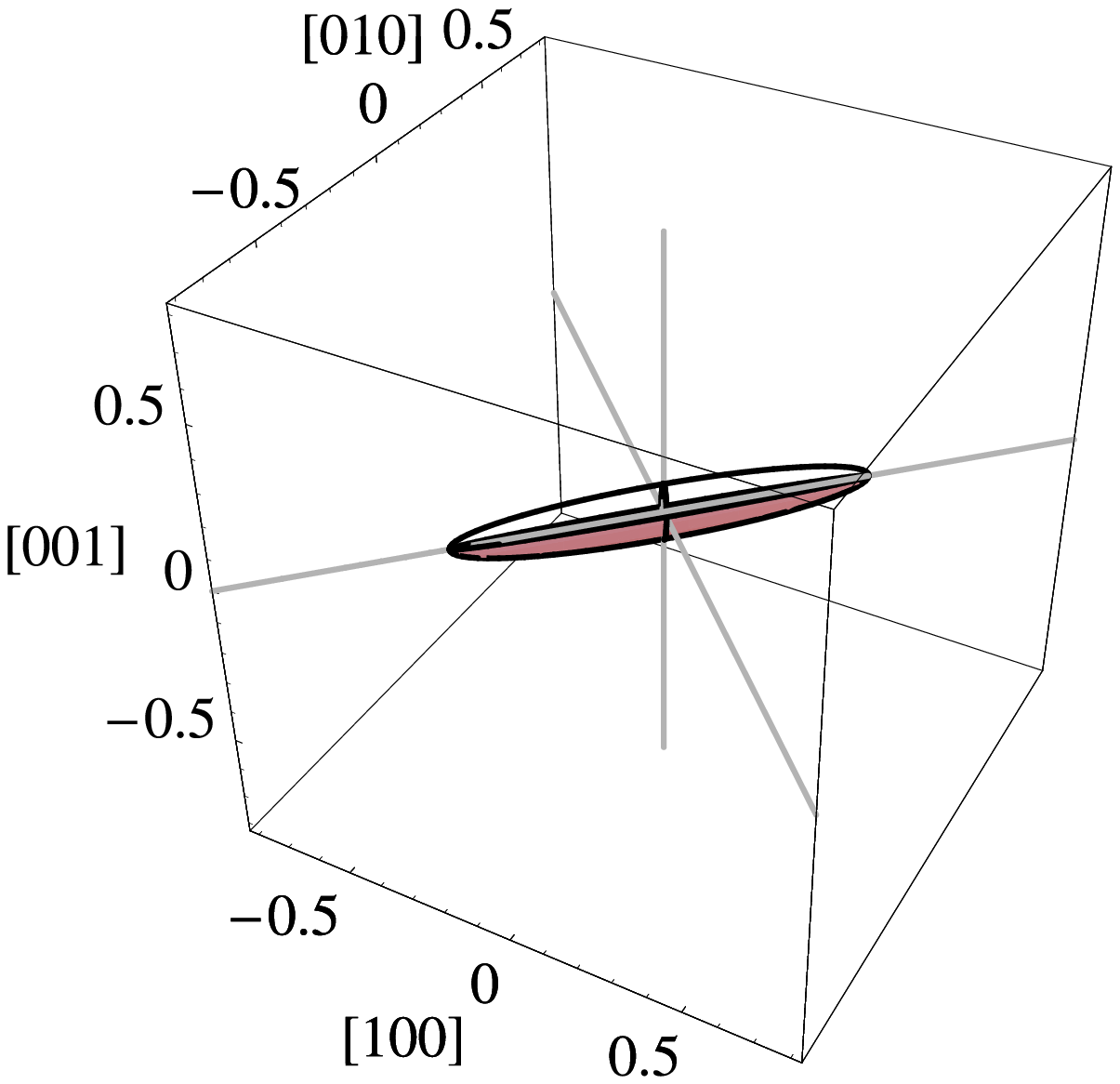}}
  \subfigure[$V_B^0$]{\label{fig:VB0vrelxxyyzz}
    \includegraphics[width=0.19\textwidth]{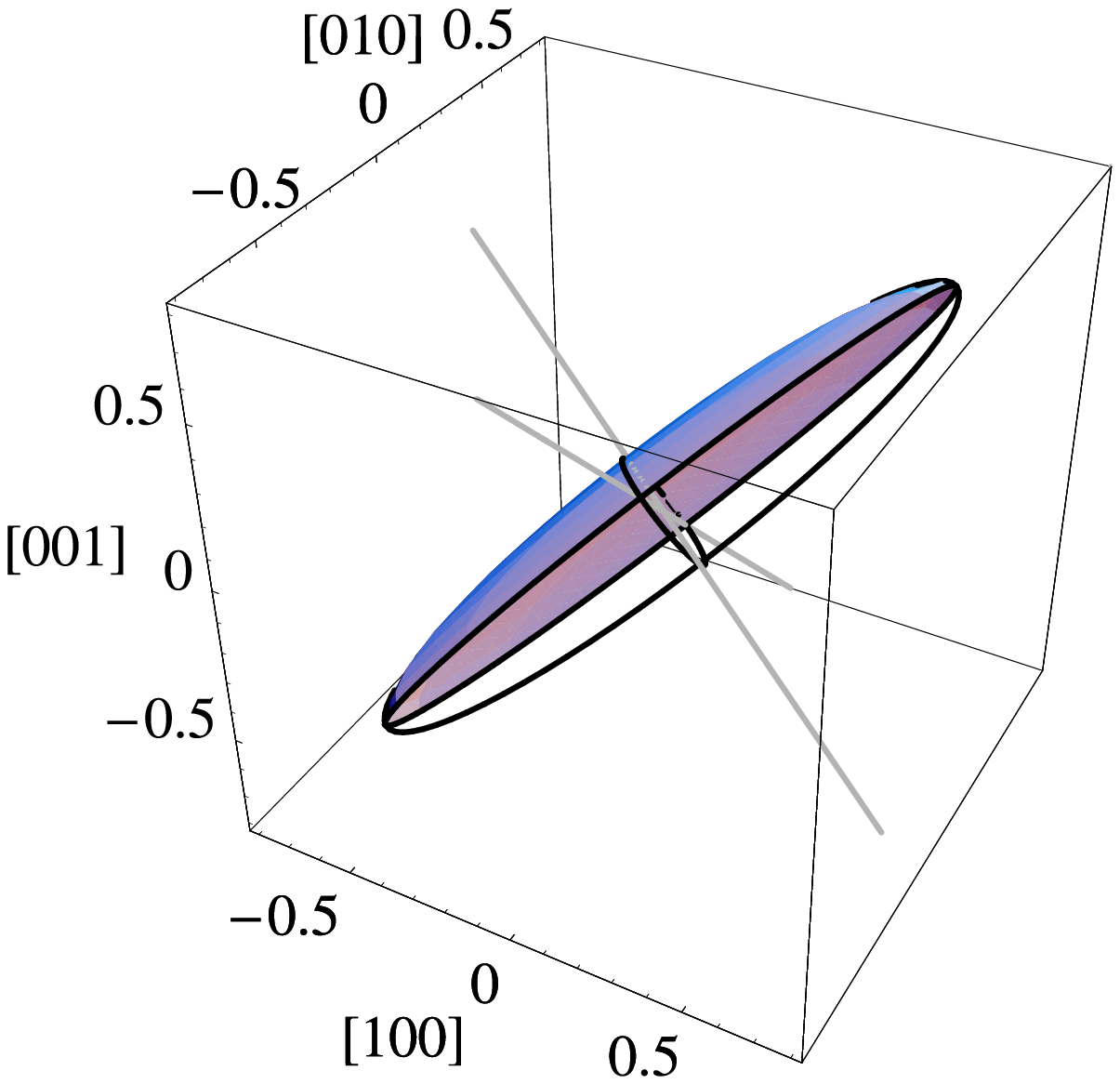}}
  \subfigure[$I_H^0$]{\label{fig:IH0vrelxxyyzz}
    \includegraphics[width=0.19\textwidth]{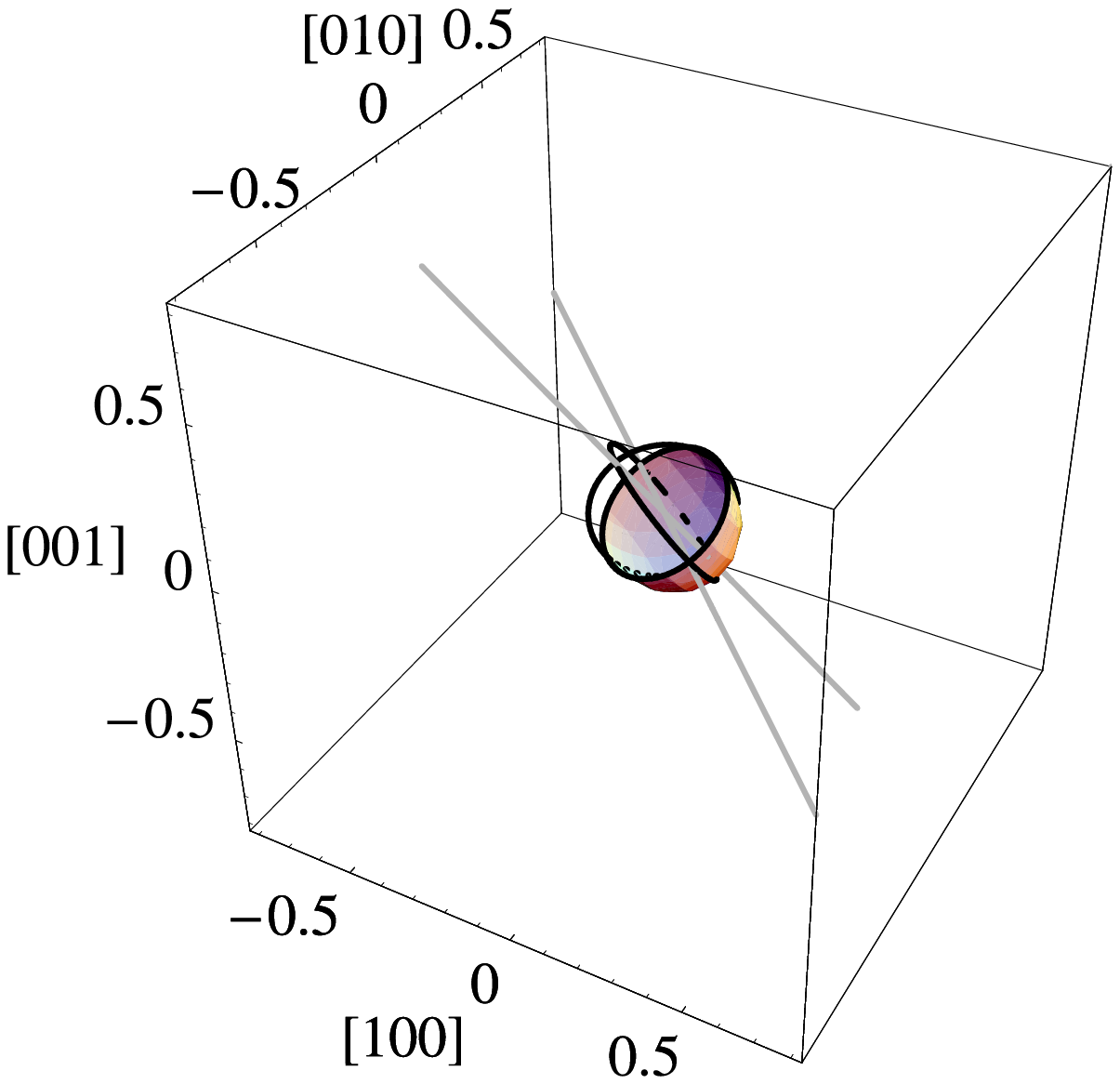}}
  \subfigure[$I_T^0$]{\label{fig:IT0vrelxxyyzz}
    \includegraphics[width=0.19\textwidth]{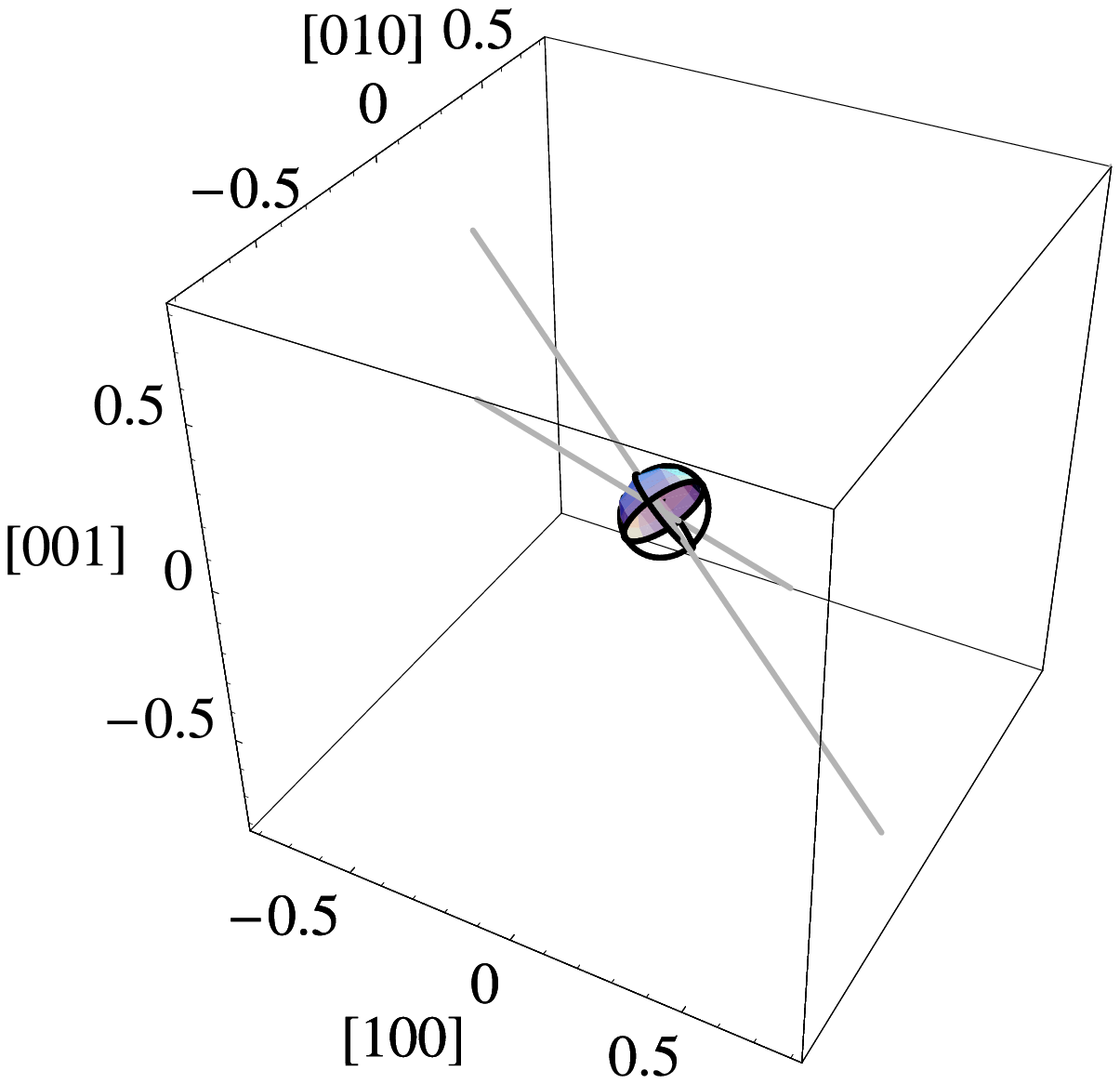}}
  \mycap{\label{fig:vrelxxyyzz}Volume ellipsoids of neutral
    defects in Si}{at $\sigma_{ij} = 0$ and $\level\fermi =
    \level\intrinsic$, as defined in Eq~\eqref{eq:vrelxxyyzz}.
    Principal axes are indicated in gray.}
\end{figure*}

\begin{figure*}[tbp]
  \centering
  \subfigure[$V_L^0$]{\label{fig:VL0vrelxyz}
    \includegraphics[width=0.19\textwidth]{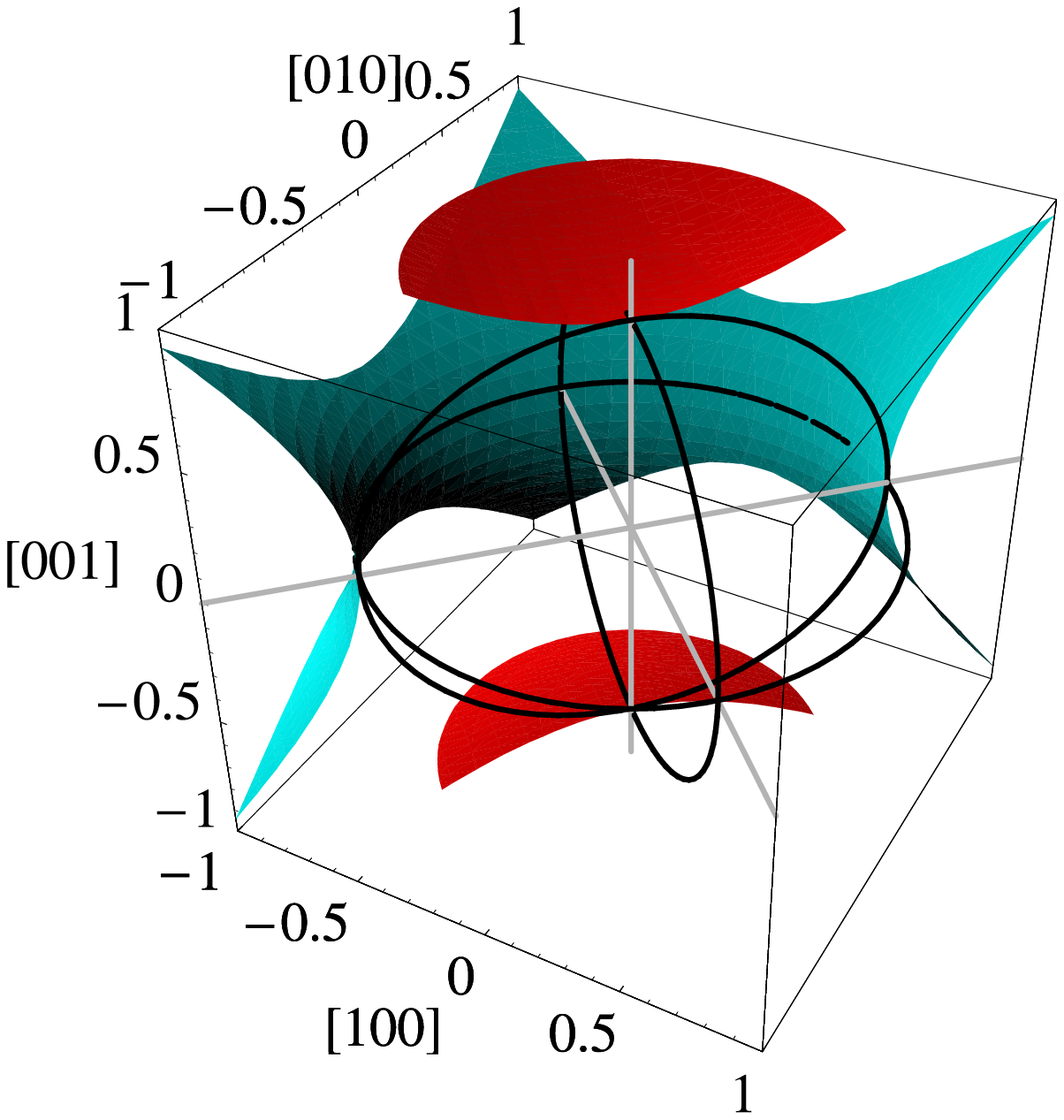}}
  \subfigure[$I_{\braket{110}}^0$]{\label{fig:IX0vrelxyz}
    \includegraphics[width=0.19\textwidth]{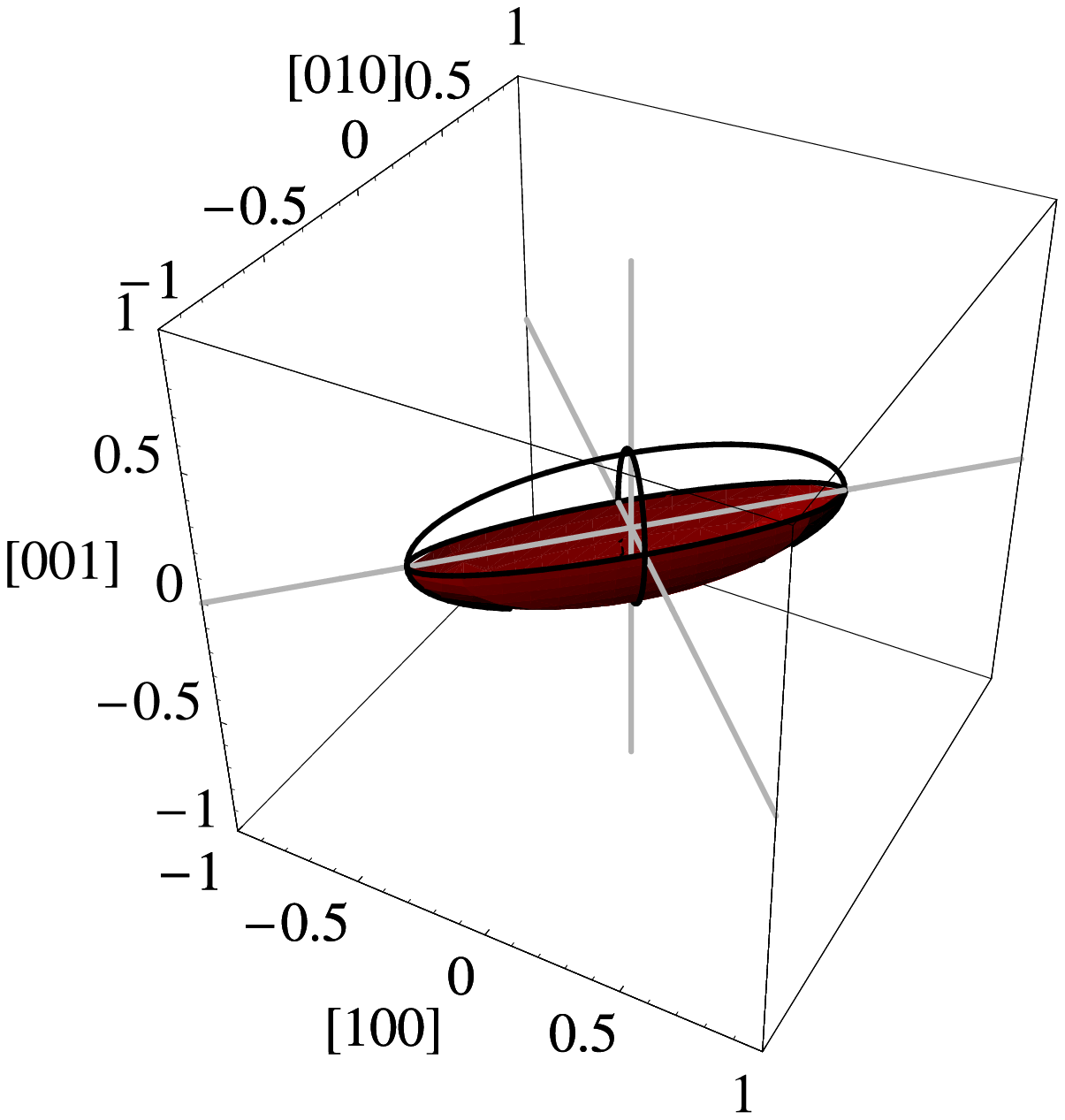}}
  \subfigure[$V_B^0$]{\label{fig:VB0vrelxyz}
    \includegraphics[width=0.19\textwidth]{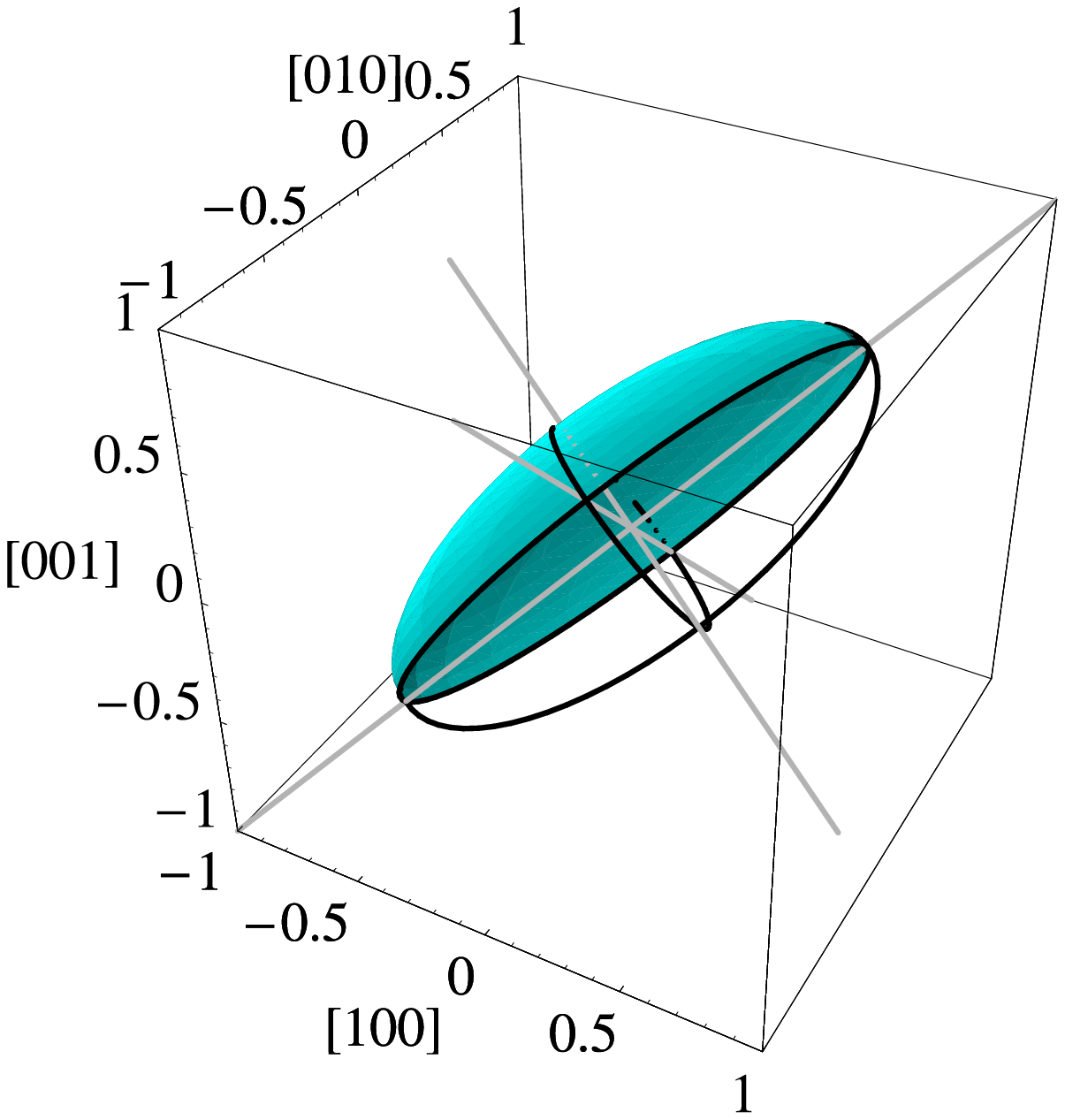}}
  \subfigure[$I_H^0$]{\label{fig:IH0vrelxyz}
    \includegraphics[width=0.19\textwidth]{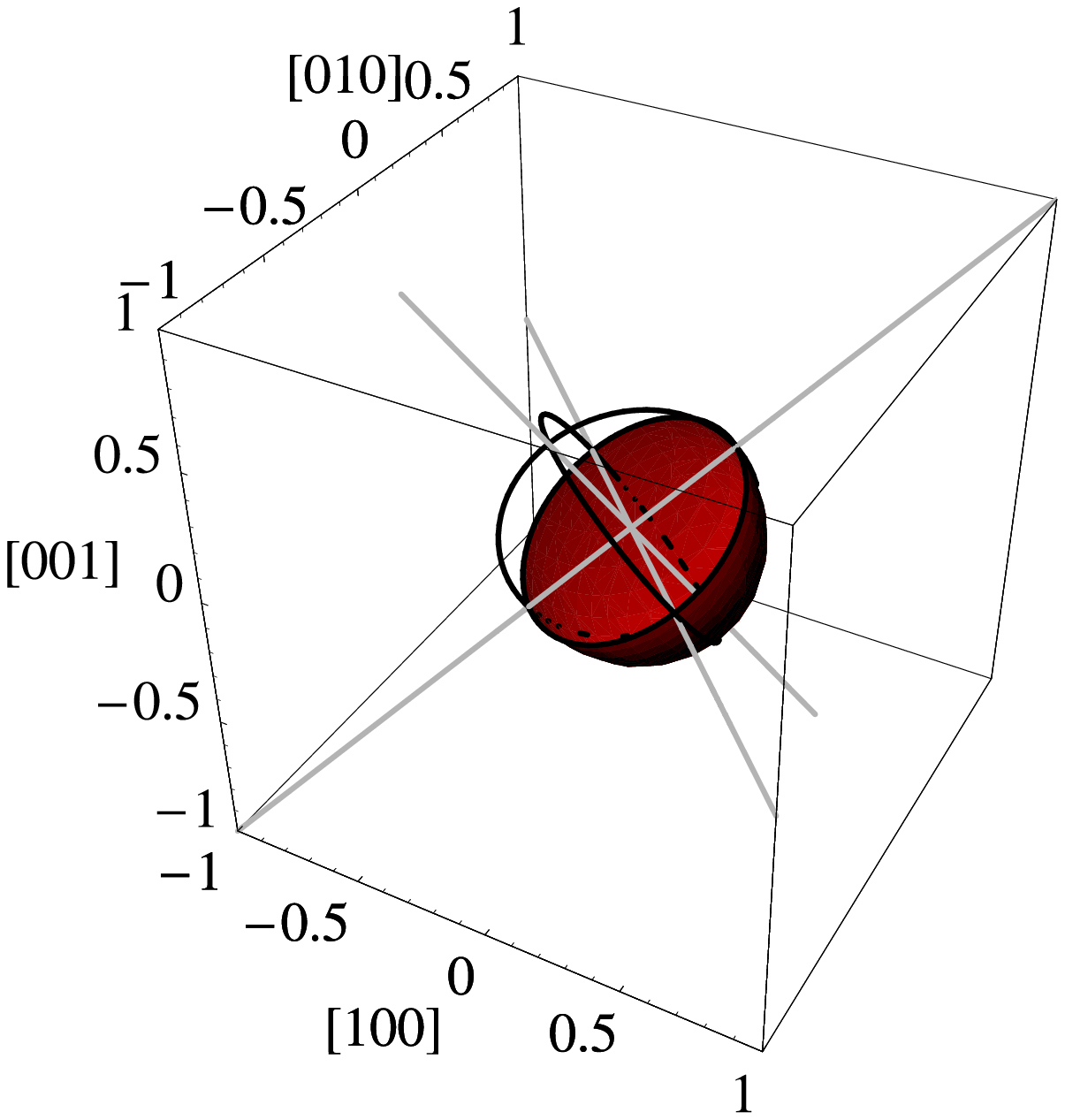}}
  \subfigure[$I_T^0$]{\label{fig:IT0vrelxyz}
    \includegraphics[width=0.19\textwidth]{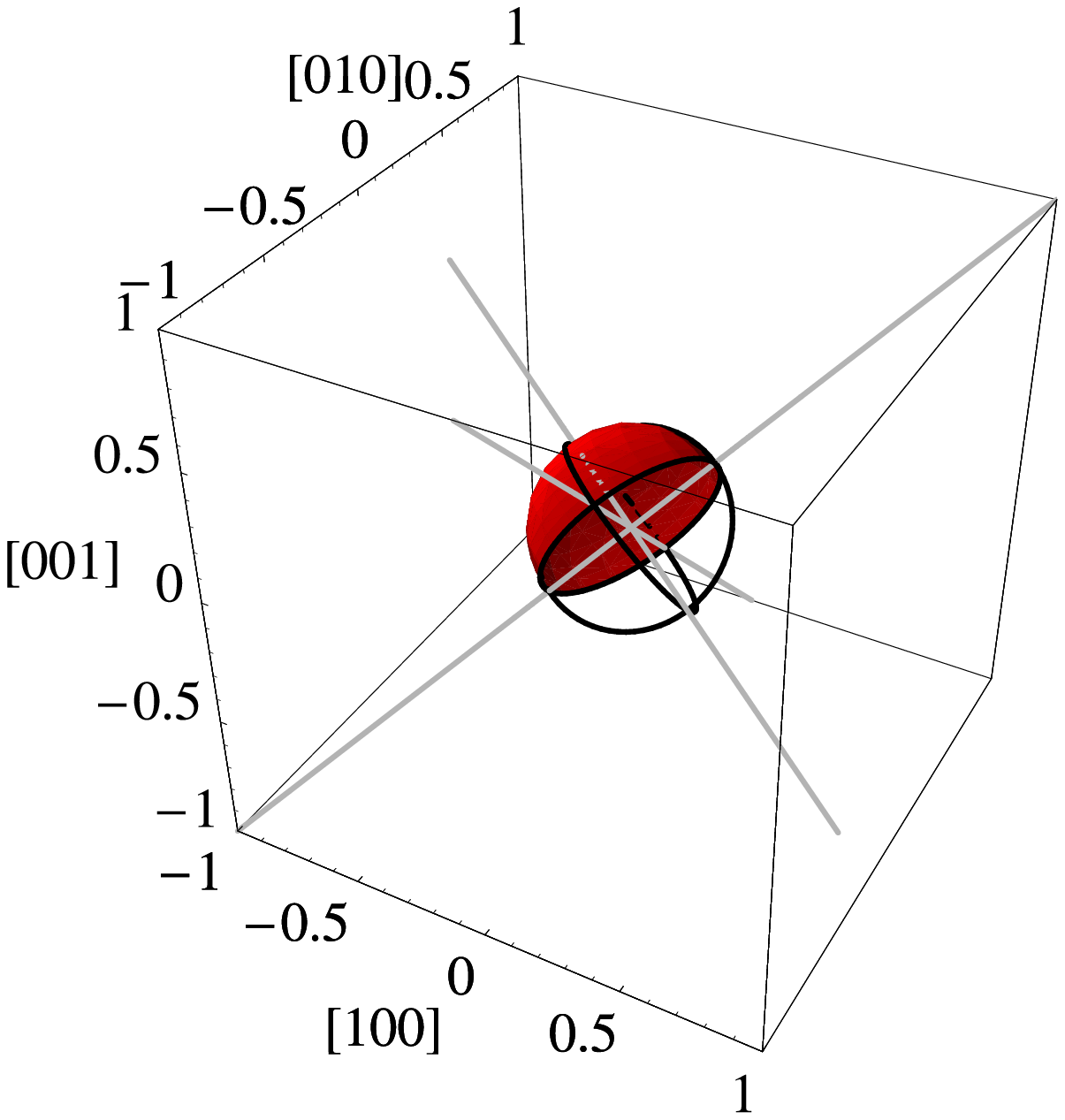}}
  \mycap{\label{fig:vrelxyz}Volume director surfaces of neutral
    defects in Si}{at $\sigma_{ij} = 0$ and $\level\fermi =
    \level\intrinsic$, as defined in Eq~\eqref{eq:vrelxyz}.
    Surfaces of both expansion (\emph{light blue}) and contraction
    (\emph{red}) are shown.  Principal axes are indicated in
    gray.}
\end{figure*}

Using \eqref{eq:relaxtens}, we have calculated the full tensor
volumes of relaxation of intrinsic defects in Si, which are mostly
quite anisotropic.  The results are more easily interpreted by
diagonalizing the matrices to find the eigenvalues and principal
directions, which we have compiled in Table~\ref{tab:relaxtens}.
(Some of the defects have an axis of symmetry, leading to
degeneracy of eigenvalues.  As a result, the corresponding
eigenvectors are not unique, and a different pair of directions in
that plane could be chosen instead.)  Useful methods of
visualizing these tensors \cite{Timoshenko51} include plotting the
volume ellipsoids (Figure~\ref{fig:vrelxxyyzz}) defined by
\begin{equation}
\label{eq:vrelxxyyzz}
  \frac{x^2}{v_1^2} + \frac{y^2}{v_2^2} + \frac{z^2}{v_3^2} = 1
\end{equation}
and the volume director surfaces (Figure~\ref{fig:vrelxyz})
\begin{equation}
\label{eq:vrelxyz}
  \frac{x^2}{v_1} + \frac{y^2}{v_2} + \frac{z^2}{v_3} = \pm 1
\end{equation}
where $+1$ is taken for expansion and $-1$ for contraction.

The sum of the eigenvalues---the scalar part of the tensor--is
approximately equal to the scalar relaxation volume calculated
previously.  The principal directions and the signs of the eigenvalues
are all in accord with intuition.  We find that the neutral lattice
vacancy has $D_{2d}$ symmetry (at least at this pressure).  The
expansion in the $[001]$ direction makes its slightly negative scalar
relaxation volume all the more noteworthy.  The degree of anisotropy
makes clear that scalar relaxation volumes are far from a complete
picture of the defect displacement field: The split vacancy has a
quite different $C_{3v}$ symmetry, drawing all six of its neighbors
in, mostly toward each other.

Among the self-interstitials, the $I_T^0$ is isotropic, and the
$I_H^0$ is nearly so, pushing out a bit more in the $(111)$ plane.
The greatest component of the displacement around $I_{\braket{110}}^0$
is, as expected, an expansion in the $[110]$ direction, but the other
components are quite small.  Two such interstitials, aligned in the
same direction, will tend to repel if their axes are parallel to the
line between them, and attract if the axes are perpendicular to the
line between them.  This suggests that clusters of these interstitials
may agglomerate in a $\{110\}$ plane with their axes aligned to
maximize their attraction.  This long-range interaction may explain
how interstitials are drawn together to form extended defects,
including $\braket{110}$ chains and eventually $\{311\}$ defects, as
seems to be the case \cite{Kim97}.

\section{Previous work}
\label{sec:previous-work}
First-principles calculation of the effect on intrinsic defect
formation energies of changing lattice parameter was performed as
early as 1984 by Car \etal{} \cite{Car84} and LDA in 1989 by
Antonelli and Bernholc \cite{Antonelli89}, using supercells with
32 atoms.  Those works did not include $I_{\braket{110}}$.  Sugino
and Oshiyama \cite{Sugino92} studied the effect of pressure on the
diffusion of group V dopants (P, As, Sb) in silicon, but not
silicon self-diffusion.  Most published reports are limited to
examining energy differences with a fixed volume.

Previous work by Zhu \cite{Zhu97} and others \cite{Lee98} used the
local density approximation (LDA), which does not include the gradient
correction of GGA (generalized gradient approximation) methods such as
PW91.  GGA methods tend to predict energies of localized states (such
as defects) with less error than LDA.  Also, those calculations were
limited to 64-site cells.  However, it has been found necessary to use
supercells containing over 200 atoms to stabilize the Jahn--Teller
distortion of $V_L^0$ (cf. Puska \etal{} \cite{Puska98}).

\section{Conclusions}
\label{sec:conclusions}
We have performed first-principles calculations on a number of
basic properties of intrinsic defects in silicon, some novel, some
not presented together in a unified analysis before.  The
relaxation volumes of electrons and holes ($+0.68 \vol$
and $-0.79 \vol$, respectively) are an appreciable
fraction of an atomic volume in magnitude.  The formation enthalpy
of a neutral vacancy is \unit{3.69}{\eV} and its migration
enthalpy is \unit{0.27}{\eV}.  The relaxation volume of the
neutral vacancy, $-1.07 \vol$, is of the expected sign but
the magnitude of the number is rather large, resulting in a
formation volume of $-0.07 \vol$, with an activation
volume of migration of $-0.24 \vol$.  That is, hydrostatic
pressure should lead to a slight \emph{increase} in equilibrium vacancy
concentration and an increase in vacancy diffusion.  The most
stable self-interstitial species, $I_T^{++}$, has a formation
enthalpy \unit{3.68}{\eV}, equal to the formation enthalpy of
$V_L^0$ within the accuracy of our method, and a formation volume
of $-0.57 \vol$ ($v\relaxation = +0.43 \vol$).

To estimate the concentration of a defect requires calculation of its
vibrational and configurational entropy.  The calculations we have
performed do not provide this information, but the near equality of
the formation enthalpies of the most stable vacancy and
self-interstitial demand that the ratio $C_I/C_V$ of their equilibrium
concentrations should not vary much with temperature.  The ratio will
depend upon pressure, however.  The formation volume of a Frenkel pair
with no clustering or amorphization should be a sizable negative
number, $-0.48 \vol$ to $-0.68 \vol$, unless interactions at high
vacancy concentrations destabilize the Jahn--Teller distortion.  These
parameters should be of interest in studies of silicon under large
elastic stresses.  They imply that increasing hydrostatic pressure
increases the equilibrium concentration of both vacancies and
self-interstitials, though the effect on vacancy concentration should
be weak.  Increasing pressure should also increase the mobility of
vacancies.

We have presented stability diagrams (akin to phase diagrams) of the
intrinsic defects in silicon, showing the complexity of vacancies
under pressure, in stark contrast to the behavior of
self-interstitials.  We have verified Bourgoin's prediction of a
crossing in the vacancy potential energy surfaces.

Finally, we have calculated the full tensor relaxation volumes of
these intrinsic defects, enabling researchers to model the biased
diffusion of both vacancies and self-interstitials under the
non-hydrostatic stress states found in actual devices.

\begin{acknowledgments}
  \auspices
\end{acknowledgments}

\bibliography{silicon}

\end{document}